\documentclass[aps,twocolumn,prb,amsmath,amssymb,superscriptaddress,floats]{revtex4}

\usepackage{amsmath}	
\usepackage{gensymb}
\usepackage{hyperref}
\hypersetup{colorlinks=true}
\usepackage{upgreek}
\usepackage{graphics}
\usepackage{hyperref}
\usepackage{epsfig}
\usepackage{color}
\usepackage{bm}
\usepackage{float}
\usepackage{ulem}               
\usepackage{dsfont}             
\usepackage{lettrine}           

\usepackage{CJK}
\usepackage{url}
\usepackage{multirow}

\graphicspath{{./}{./Figures/}}

\newcommand{\ii}{\imath}   
\newcommand{\diag}[1]{\mathrm{diag}\left(#1\right)}

\let\oldsection\subsection      

\usepackage{titlesec}
\titleformat*{\section}{\normalfont\normalsize\bfseries\sffamily}
\titleformat{\subsection}[runin]
  {\normalfont\normalsize\bfseries}{\thesubsection}{1em}{}
\titlespacing*{\section}
  {0pt}{5.5ex plus 1ex minus .2ex}{0.3ex plus .2ex}

\begin{document}

\title{Gapless quantum spin liquid in a honeycomb $\Gamma$ magnet}
\author{Qiang Luo}
\affiliation{Department of Physics, Renmin University of China, Beijing 100872, China}
\affiliation{Department of Physics, University of Toronto, Toronto, Ontario M5S 1A7, Canada}
\author{Jize Zhao}
\email[]{zhaojz@lzu.edu.cn}
\affiliation{School of Physical Science and Technology $\&$ Key Laboratory for Magnetism and
Magnetic Materials of the MoE, Lanzhou University, Lanzhou 730000, China}
\author{Hae-Young Kee}
\affiliation{Department of Physics, University of Toronto, Toronto, Ontario M5S 1A7, Canada}
\affiliation{Canadian Institute for Advanced Research, Toronto, Ontario, M5G 1Z8, Canada}
\author{Xiaoqun Wang}
\email[]{xiaoqunwang@sjtu.edu.cn}
\affiliation{Key Laboratory of Artificial Structures and Quantum Control (Ministry of Education), Shenyang National Laboratory for Materials Science, School of Physics and Astronomy, Tsung-Dao Lee Institute, Shanghai Jiao Tong University, Shanghai 200240, China}
\affiliation{Beijing Computational Science Research Center, Beijing 100084, China}

\date{\today}

\begin{abstract}
    A family of spin-orbit coupled honeycomb Mott insulators offers a playground to search for quantum spin liquids (QSLs) via bond-dependent interactions. In candidate materials, a symmetric off-diagonal $\Gamma$ term, close cousin of Kitaev interaction, has emerged as another source of frustration that is essential for complete understanding of these systems. However, the ground state of honeycomb $\Gamma$ model remains elusive, with a suggested zigzag magnetic order. Here we attempt to resolve the puzzle by perturbing the $\Gamma$ region with a staggered Heisenberg interaction which favours the zigzag ordering. Despite such favour, we find a wide disordered region inclusive of the $\Gamma$ limit in the phase diagram. Further, this phase exhibits a vanishing energy gap, a collapse of excitation spectrum, and a logarithmic entanglement entropy scaling on long cylinders, indicating a gapless QSL. Other quantities such as plaquette-plaquette correlation are also discussed.
\end{abstract}

\pacs{}

\maketitle

\section*{INTRODUCTION}

The ongoing search for exotic magnetic states in highly frustrated antiferromagnets
\cite{Balents2010,HanHCetal2012,LiYMGO2015,LiaoXiang2017,WangSandvik2018,HuZEH2019}
has been extended to a new class of correlated materials
with a two-dimensional honeycomb structure\cite{PlumbCSetal2014,KimKee2015,BanerjeeNatMat2016,LiYangZhouetal2019}
and its three-dimensional variants\cite{RauLeeKee2016}.
It is suggested that bond-dependent interactions could be realized in the spin-orbit coupled Mott insulators
with the aforementioned lattice geometry\cite{Jackeli2009}.
In particular, the Kitaev honeycomb model exhibits a novel Kitaev quantum spin liquid (QSL)
which hosts fractionalized Majorana fermions and flux excitations\cite{Kitaev2006}.
Realization of Kitaev interaction in real materials was first proposed in iridates
\cite{YeCCetal2012,ChoiCKetal2012,ChunKKetal2015},
and then turned toward $\alpha$-RuCl$_3$ in which
Ru$^{3+}$ ions are arranged in a honeycomb lattice and carry effective spin-$1/2$ particles
\cite{PlumbCSetal2014,BanerjeeNatMat2016}.
Although $\alpha$-RuCl$_3$ displays long-range zigzag magnetic order at low temperature
\cite{SearsSongPlumbetal2015,LeahyPRL2017,SearsZhaoLynnetal2017,BaekPRL2017,Wolter2017,ZhengWenYu2017},
it is argued to be proximate to the Kitaev QSL owing to the broad continuum of magnetic excitations identified in
Raman scattering\cite{SandTianPluetal2015,YeMeiHuangGroupEtal2019}
and inelastic neutron scattering\cite{BanerjeeNatMat2016,RanYuLiWen2017,DoNatPhys2017,WinterNcom2018}.

In spite of massive research efforts, it has been challenging to determine exchange parameters
of the proposed spin Hamiltonian for $\alpha$-RuCl$_3$
(see refs.~\cite{JanAndVoj2017,LaurellOkam2020} and references therein).
However, there is a broad consensus on a sizable off-diagonal $\Gamma$ interaction\cite{RanLeeKeePRL2014,YadavSciRep2016}
which is antiferromagnetic (AFM) and is potentially comparable to the celebrated Kitaev interaction\cite{RanYuLiWen2017,WangDYL2017,GordonCSetal2019}.
Crucially, it is shown that the $\Gamma$ interaction could help enhance the mass gap of Majorana fermions\cite{TakikawaFujimoto2019}
and is responsible for the strongly anisotropic responses to the magnetic field observed in $\alpha$-RuCl$_3$
provided that the Land{\'{e}} $g$-factor anisotropy is modest\cite{RanLeeKeePRL2014,JanAndVoj2017,Lampen-KelleyArxiv2018}.
In contrast to the Kitaev model\cite{Kitaev2006},
analytical solution of the honeycomb $\Gamma$ model has not been found yet\cite{SamWachYamaetal2018}.
Previous classical studies have demonstrated that its ground state is a classical spin liquid\cite{RousochatzakisPerkins2017},
followed by a flux-ordered spin liquid which is stabilized in a finite temperature window\cite{SahaFZetal2019}.
Given the infinite classical ground-state degeneracy\cite{RousochatzakisPerkins2017},
determining the precise quantum nature of $\Gamma$ model is nontrivial,
and existing numerical works have already led to conflicting results.
Parallel works by exact diagonalization\cite{CatunYWetal2018} 
and density-matrix renormalization group~(DMRG) study of a cylinder with a width of three unit cells\cite{GohlkeWYetal2018}
both claim that the ground state is a nonmagnetic phase.
A variational Monte Carlo simulation, on the other hand, suggests that it is a zigzag order\cite{WangBLArxiv2019}.
Furthermore, a recent study proposes that it is a nematic paramagnet that spontaneously breaks the lattice rotational symmetry\cite{GohlkeCKK2020}.

In this work, we study a model which consists of the $\Gamma$ term and of a staggered Heisenberg ($\tilde{J}$) interaction
along the bonds, dubbed the \textit{bond-modulated} $\tilde{J}$-$\Gamma$ model (see Eq.~\eqref{BMJG-Ham}).
Depending on the sign of $\tilde{J}$, it could either favor the zigzag order ($\tilde{J} > 0$) or stripy order ($\tilde{J} < 0$).
If the ground state of $\Gamma$ model is a zigzag ordered phase, 
then the zigzag order protruding from the pure $\Gamma$ limit should compete with
and survive up to a finite ferromagnetic $\tilde{J}$ interaction.
Otherwise, there will be an intermediate phase sandwiched between the two magnetically ordered states.
Thus, this model works as a virtuous arena to clarify the debates by unfolding the competing states,
although it is not a description of any particular material.
By employing the DMRG method on both finite cylinders with circumferences of up to 10 sites
and $C_3$-symmetric hexagonal clusters\cite{White_1992,StoudenmireWhite_2012},
we identify a disordered state in between.
This phase manifests characters of a gapless QSL including a dense excitation spectrum,
logarithmic entanglement entropy scaling, and short-range plaquette-plaquette correlation.
The pure $\Gamma$ limit belongs to this QSL and is separated from the zigzag order by a first-order transition.


\section*{RESULTS}

\subsection*{Model\\}
\begin{figure}[!ht]
\centering
\includegraphics[width=0.95\columnwidth, clip]{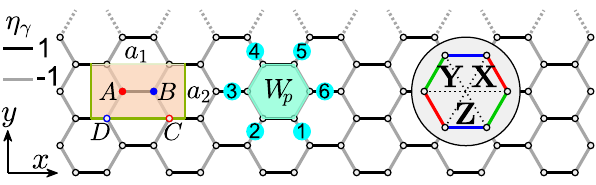}\\
\caption{\textbf{Lattice geometry.}
    Illustration of an XC6 cylinder on a honeycomb lattice.
    $\eta_{\gamma}$ is $+1$~($-1$) for horizontal~(zigzag) bonds.
    The insets are (left) the unit cell for the zigzag/stripy order with $a_1 = 3$ and $a_2 = \sqrt3$,
    (middle) the hexagonal plaquette operator $\hat{W}_p$ with its six sites enumerated,
    and (right) the $\textbf{X}$~(red), $\textbf{Y}$~(green), and $\textbf{Z}$~(blue) bonds.}\label{FIG-XC6}
\end{figure}

The Hamiltonian of the bond-modulated $\tilde{J}$-$\Gamma$ model reads
\begin{align}\label{BMJG-Ham}
\mathcal{H} =& \tilde{J} \sum_{\left<ij\right>\parallel{\gamma}} \eta_{\gamma}\textbf{S}_i\cdot\textbf{S}_j
+ \Gamma \sum_{\left<ij\right>\parallel\gamma}(S_i^{\alpha}S_j^{\beta}+S_i^{\beta}S_j^{\alpha})
\end{align}
where $S_i^{\gamma}$~($\gamma$ = $x$, $y$, $z$) is the $\gamma$-component of a spin-1/2 operator at site $i$,
and $\alpha$ and $\beta$ are the two other bonds on a honeycomb lattice.
$\eta_{\gamma} = 1$ for the bond $\langle ij\rangle_{\gamma}$ along the horizontal direction
and equals to $-1$ otherwise~(see Fig.~\ref{FIG-XC6}).
$\tilde{J}$ and $\Gamma$ are parameterized using $\vartheta\in[0,\pi]$ so as to
$\tilde{J} = \cos\vartheta$ and $\Gamma = \sin\vartheta~(\geq 0)$.

In what follows, we carry out a hierarchical study of Eq.~\eqref{BMJG-Ham}
to provide multi-faceted evidences of the gapless QSL nature of $\Gamma$ magnet.
We start by mapping out the classical phase diagram via the parallel tempering Monte Carlo simulation\cite{Metropolis1953,HukushimaNemoto1996},
and conclude that the ground state of $\Gamma$ model sits exactly at the classical transition point on the verge of the zigzag phase.
This makes sense because the zigzag ordering belongs to the macroscopic ground-state manifold of the classical $\Gamma$ model\cite{RousochatzakisPerkins2017}.
Next, we show the energy reduction and sublattice magnetization within the linear spin-wave analysis (for a review, see Ref.\cite{JanssenVojta2019}).
Afterwards, we present a quantum phase diagram obtained by large-scale DMRG calculations on various distinct cluster geometries.

\begin{figure}[!ht]
\centering
\includegraphics[width=0.95\columnwidth, clip]{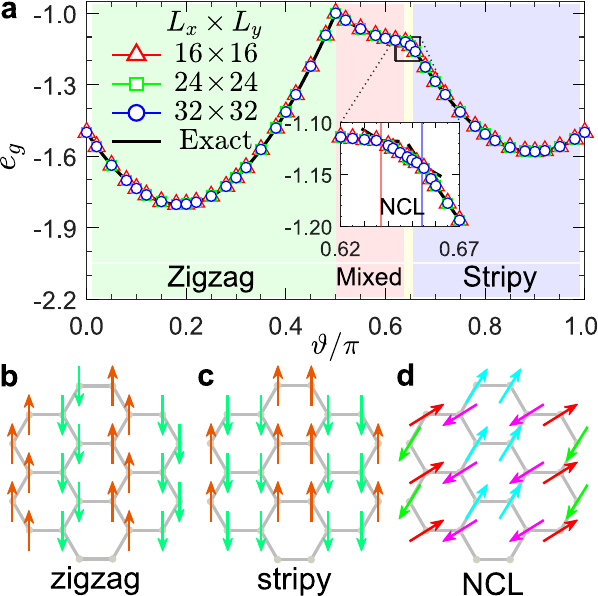}\\
\caption{\textbf{Classical phase diagram.}
    \textsf{\textbf{a}} Monte Carlo simulations of classical energy $e_g$ under three XC clusters
    of $16\times16$~(red triangle), $24\times24$~(green square), and $32\times32$~(blue circle).
    The solid black line stands for the exact solution given by the energy optimization method.
    The classical phase diagram is then determined by kinks in energy curves.
    Inset: Zoom in of energy curves near $\vartheta_{t,r}^{\textrm{cl}}/\pi \approx 0.6476$.
    A noncollinear (NCL) phase appears in a narrow window of $0.6368 < \vartheta/\pi < 0.6543$.
    \textsf{\textbf{b}} and \textsf{\textbf{c}} depict the zigzag order and stripy order, respectively.
    \textsf{\textbf{d}} A NCL phase with a unit cell of $4\times2$.
    }\label{FIG-ClEg}
\end{figure}

\begin{figure*}[htpb!]
\centering
\includegraphics[width=1.00\linewidth, clip]{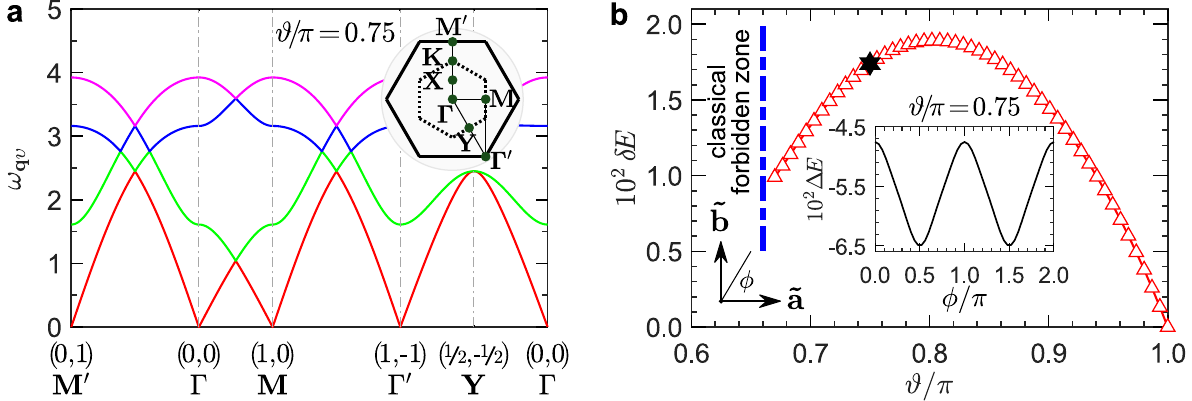}\\
\caption{\textbf{Spin-wave analysis of the stripy phase.}
    \textsf{\textbf{a}} Four branches of the magnon spectra $\omega_{\textbf{q}\upsilon}$ in the stripy phase where $\vartheta/\pi = 0.75$.
    The path along the symmetry directions in the momentum space is depicted in the inset.
    \textsf{\textbf{b}} Energy barrier $\delta E$ between the stripy phases of different orientations in the classically allowed zone.
    Inset: Quantum energy correction $\Delta E(\phi)$ vs angle $\phi$ suited at the $\mathbf{\tilde a}$-$\mathbf{\tilde b}$ plane.
    The parameter is fixed to $\vartheta/\pi = 0.75$, which is marked as a black hexagram in the main panel.
    }\label{FIG-LSWT}
\end{figure*}

\subsection*{Classical phase diagram\\}
Figure~\ref{FIG-ClEg} shows the classical phase diagram of Eq.~\eqref{BMJG-Ham} obtained by Monte Carlo simulations\cite{Metropolis1953,HukushimaNemoto1996},
coincide exactly with the subsequent results of energy optimization method (see ``Methods'' section).
Due to the bond-modulated $\eta_{\gamma}$-term, the conventional zigzag and stripy orderings perpendicular to the $\textbf{Z}$ bonds
are induced when $\vartheta/\pi$ is 0 or 1, respectively.
By introducing AFM $\Gamma$ interaction, the ground state becomes more competitive,
triggering the possibility of other magnetic orderings in the moderate interaction regime.
The ground-state energy $e_g = E_g/(NS^2)$ ($E_g$ is the total energy) is shown in Fig.~\ref{FIG-ClEg}a,
while selected spin configurations of the corresponding phases are depicted in Fig.~\ref{FIG-ClEg}b-d.
In the phase diagram, the leftmost is the zigzag order with $e_g^{\textrm{zz}} =  -(2\Gamma+3\tilde{J})/2$
and its magnetic moment direction is $\mathbf{n}\;[11\bar1]$.
The rightmost is occupied by the stripy order with $e_g^{\textrm{st}} =  -(\Gamma-3\tilde{J})/2$.
Its spins are perpendicular to $\mathbf{n}~[11\bar{1}]$, but could vary freely in the plane spanned by
$\mathbf{\tilde{a}}~[112]$ and $\mathbf{\tilde{b}}~[1\bar{1}0]$, showing an emergent continuous symmetry.
Further, an extensive intermediate region appears in between.
It is dominated by a so-called mixed phase in which the AFM order and two twining zigzag orders are degenerate
with energy $e_g^{\textrm{mixed}} =  -(2\Gamma-\tilde{J})/2$.
Here, twining zigzag orders refer to the other two zigzag orders whose spin orientations
are different from the one shown in Fig.~\ref{FIG-ClEg}b
(for spin configurations, see \underline{Supplementary Note 1}).
The zigzag--mixed transition takes place exactly at $\vartheta_{t,l}^{\textrm{cl}}/\pi = 0.5$,
reflecting the classical spin liquid of the $\Gamma$ model\cite{RousochatzakisPerkins2017}.
There is no direct transition between the mixed phase and the stripy phase
expected to occur at $\vartheta_{t,r}^{\textrm{cl}}/\pi$ = $1-\frac{1}{\pi}\textrm{atan}\,2$ $\approx 0.6476$.
Instead, a noncollinear phase (see Fig.~\ref{FIG-ClEg}d) with $e_g$ =  $-\sqrt{\tilde{J}^2+\Gamma^2/16}$ $-$ $\Gamma/\sqrt2$
appears in a narrow window of $\vartheta/\pi$ that is less than 0.02, see inset of Fig.~\ref{FIG-ClEg}a.

\subsection*{Spin-wave theory\\}
To understand the role played by quantum fluctuations and for the sake of comparison with the DMRG results later,
we have performed the linear spin-wave calculation\cite{JanssenVojta2019}
based on the quadratic Hamiltonian
$\bar{\mathcal{H}} = E_g[S^2\to S(S+1)] + \frac{S}{2} \sum_{\bm{q}} \psi_{\bm q}^{\dagger} \mathcal{M}_{\bm q}\psi_{\bm q}$,
where $\psi_{\bm q}^{\dagger} = \big(a_{\bm q}^{\dagger}, b_{\bm q}^{\dagger}, \cdots, a_{-{\bm{q}}}, b_{-\bm q}, \cdots \big)$ is the Nambu spinor,
and $\mathcal{M}_{\bm q}$ is a $2\times2$ block matrix (see ``Methods" and \underline{Supplementary Note 2}).
There are four spin-wave dispersion branches $\omega_{{\bm q}\upsilon}$~($\upsilon$ = 1-4) for the four-sublattice ($n_s = 4$) zigzag and stripy orderings.
In the zigzag order, there exists a magnon gap $\Delta$ at $\mathbf{M}$ point in the Brillouin zone (see inset of Fig.~\ref{FIG-LSWT}a).
When approaching $\Gamma$ limit, $\tilde{J}/\Gamma \ll 1$,
the lowest magnon branch is softened and the gap vanishes as $\Delta/\Gamma \simeq \frac{\sqrt{30}}{3}\sqrt{\tilde{J}/\Gamma}$.
Therefore, the zigzag order could only survive for AFM $\tilde{J}$ (i.e., $\vartheta/\pi < 0.50$),
beyond which the magnon branch becomes imaginary and should be terminated by a transition.

\begin{figure}[!ht]
\centering
\includegraphics[width=1.00\linewidth, clip]{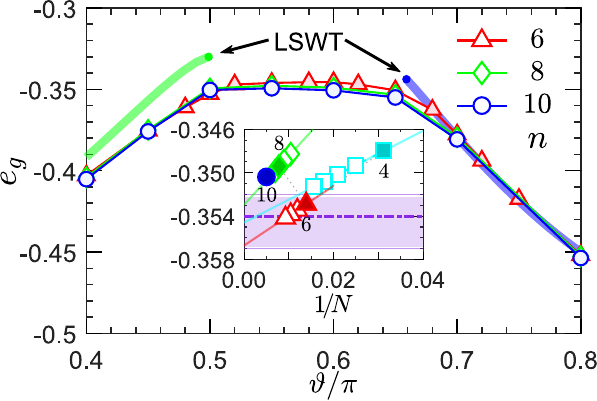}\\
\caption{\textbf{Quantum ground-state energy.}
    DMRG result of $e_g$ under three XC clusters where the circumferences $n$ are 6~(red triangle), 8~(green diamond), and 10~(blue circle).
    The thick belts are the energy of the zigzag order (green belt) and stripy order (blue belt) obtained by the linear spin-wave theory (LSWT).
    Inset: Extrapolation of energy for $\Gamma$ model. For each circumferences $n$ the energy is linearly decreasing with $1/N$ and the special cases ($L_x/L_y = 2$) are marked by filled symbols. The extrapolated values fall in the purple band centered at $-0.354(3)$.}\label{FIG-EgTht}
\end{figure}

\begin{figure*}[htpb!]
\centering
\includegraphics[width=0.80\linewidth, clip]{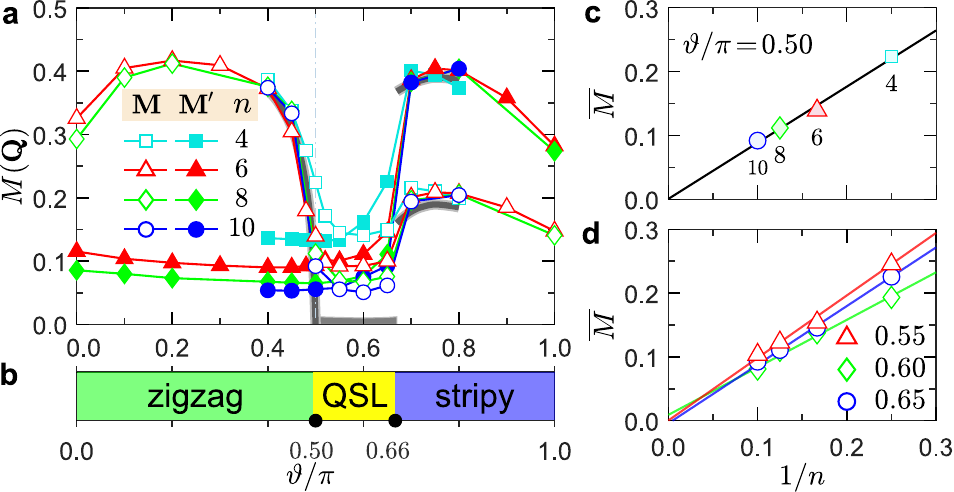}\\
\caption{\textbf{Magnetic order parameters and quantum phase diagram.}
    \textsf{\textbf{a}} Magnetic order parameters $M(\textbf{Q})$ for the zigzag order (open symbols) and stripy order (filled symbols)
    with $\textbf{Q} = \textrm{\bf{M}}$ and/or $\textrm{\bf{M}}'$ under four finite XC clusters.
    The thick gray line shows the magnetic order in the thermodynamic limit.
    \textsf{\textbf{b}} Quantum phase diagram of the bond-modulated $\tilde{J}$-$\Gamma$ model.
    \textsf{\textbf{c}} and \textsf{\textbf{d}} show extrapolations of the maximal peaks $\overline{M}$ throughout the reciprocal space.
    \textsf{\textbf{c}} Linearly extrapolation for $\Gamma$ model with $\vartheta/\pi = 0.50$.
    \textsf{\textbf{d}} $\vartheta/\pi$ = 0.55 (red triangle), 0.60 (green diamond), and 0.65 (blue circle), respectively.}\label{FIG-MSFTht}
\end{figure*}

The magnon spectra for the representative stripy order with $\vartheta/\pi = 0.75$ are shown in Fig.~\ref{FIG-LSWT}a
where the wave vector $\bm q$ is parameterized in units of $(h, k)$ as
$\bm q$ = $\big(\frac{2\pi}{a_1}h, \frac{2\pi}{a_2}k\big)$\cite{ChoiCKetal2012}.
The spectra are symmetric with the middle of the $\boldsymbol\Gamma$-$\mathbf{M}$ line,
so the $\boldsymbol\Gamma$ and $\mathbf{M}$ points are equivalent.
Due to the emergent continuous symmetry of the classical stripy order, the magnon spectra are gapless.
In the presence of quantum fluctuations, however, the degeneracy is lifted via order-by-disorder mechanism\cite{Henley1989},
selecting two of them that are either parallel or antiparallel to $\mathbf{\tilde{b}}$ axis (see inset of Fig.~\ref{FIG-LSWT}b).
To illustrate it, we firstly define the quantum energy correction
$\Delta E(\phi) = S e_g^{\textrm{st}} + \frac{S}{2n_s} \sum_{\upsilon} \int \frac{d^2\mathbf q}{(2\pi)^2} \omega_{{\bm q}\upsilon}(\phi)$,
where $\phi$ is the angle in the $\mathbf{\tilde{a}}$-$\mathbf{\tilde{b}}$ plane\cite{ZhuMakWhiteetal2017}.
For $\vartheta/\pi = 0.75$, which is deep in the stripy order,
we show $\Delta E(\phi)$ vs $\phi$ in the inset of Fig.~\ref{FIG-LSWT}b.
The energy correction has its minima at $\phi = \pi/2$ or $3\pi/2$,
corresponding to the two mostly favored configurations at the quantum level.
The energy barrier $\delta E$, defined as the energy difference between $\Delta E(\pi/2)$ and $\Delta E(0)$,
is approximately 0.0175.
The main panel of Fig.~\ref{FIG-LSWT}b shows energy barrier at different $\vartheta/\pi$ in the stripy order.
When $\vartheta/\pi = 1.00$ the energy barrier is zero, consistent with the gapless Goldstone modes thereof.
Beyond that, the energy barrier is finite, indicating that the stripy order should also be twofold degenerate in the quantum case.

Due to the magnon instabilities of zigzag and stripy orderings, they could only exist in their classically allowed regions.
The corresponding phase transitions could be illuminated by their sublattice magnetization.
It is observed that magnetization of the zigzag order is almost saturated when $\vartheta/\pi < 0.4$.
As $\vartheta/\pi$ approaches $0.5$, it undergoes a considerable suppression and
the lowest branch of the magnetization nearly vanishes at $\vartheta_{t,l}^{\textrm{cl}}/\pi = 0.50$
(see \underline{Supplementary Fig.~S5}).
The stripy order is more stable and its magnetization only has a small reduction at $\vartheta_{t,r}^{\textrm{cl}}/\pi \approx 0.6476$.
However, spin-wave energy in the mixed phase, say AFM order, is overwhelmingly higher than its neighbors.
It thus implies that the genuine phase in the intermediate region should be different from its classical counterpart,
imposing restrictions on the applicability of the spin-wave analysis.

\subsection*{Intervening magnetically disordered state\\}
As discussed, the spin-wave calculation fails in the intermediate regime, hence the quantum study is necessary.
We have performed the standard DMRG computation on three XC clusters of $12\times6$ ($n = 6$), $16\times8$ ($n = 8$), and $20\times10$ ($n = 10$)
and compute the ground-state energy $e_g = E_g/N$ which is shown in Fig.~\ref{FIG-EgTht}.
The energy curves in the middle are very flat, while they have two sharp downwarping when away from the middle region,
leading to two well-marked kinks that are signals of first-order transitions.
These discontinuous phase transitions could also be advocated by the entanglement entropy which has a trend to jump
as the system size is increased (see \underline{Supplementary Fig.~S9 and Fig.~S11}).
Therefore, the DMRG result supports an intermediate region that impedes the direct transition between the zigzag and stripy phases\cite{HuangSu2016,HuangSu2017}.
For comparison, we also depict the spin-wave energy of the zigzag order (green belt) and stripy order (blue belt) in Fig.~\ref{FIG-EgTht}.
It is clearly found that there is a further energy reduction of the zigzag order beyond the linear approximation,
in accord with the dramatic suppression of magnetic order parameter which will be clarified later.
Strikingly, as shown in the inset of Fig.~\ref{FIG-EgTht},
the energy $e_g$ of $\Gamma$ model ($\vartheta/\pi = 0.5$) exhibits a nonmonotonic scaling behavior\cite{LuoKG2021},
indicative of a possible periodicity as revealed in the Kitaev model\cite{Kitaev2006}.
At each fixed circumference $n$, the energy is linearly decreasing with length $L_x$ of the cylinder.
By varying the circumference $n$ from 4 to 10, the extrapolated energy has an oscillation in a window of $-0.357 < e_g < -0.352$.
Therefore, we estimate that the energy of $\Gamma$ model is $e_g=-0.354(3)$ in the thermodynamic limit.

In order to unveil the nature of the intermediate region and to pin down the precise phase boundaries,
we resort to the magnetic order parameter,
which is defined as $M_N({\bf{Q}}) = \sqrt{\mathbb{S}_N({\bf{Q}})/N}$
where $\mathbb{S}_N({\bf{Q}})$ is the magnetic structure factor
with ${\bf{Q}}$ being the ordering wavevector (see ``Methods" for definition).
The zigzag phase has a peak at $\textbf{M}$ point in the Brillouin zone,
while the stripy phase possesses both peaks at $\textbf{M}$ and $\textbf{M}'$ points.
Crucially, magnetic structure factor of the intermediate region is diffuse with a soft peak.
Fig.~\ref{FIG-MSFTht}a displays order parameters $M_N({\bf{Q}})$ of the zigzag and stripy phases
on four distinct XC clusters with a circumference ranging from 4 to 10.
Akin to their spin-wave results, magnetization of the zigzag and stripy phases exhibits maxima at $\vartheta/\pi \approx$ 0.25 and 0.75, respectively.
This implies that these magnetic orderings are most stable when $\Gamma$ term is approximately of equal strength to the Heisenberg interaction.
Away from these points, quantum fluctuations are enhanced so that the magnetic ordering in the intermediate region is dramatically suppressed,
followed by an algebraically decay with the circumference $n$ (see Fig.~\ref{FIG-MSFTht}c and Fig.~\ref{FIG-MSFTht}d).
After a careful inspection of the finite-size effect, we conclude that the magnetization will disappear eventually
as the best fitting gives $M \to 0.0$ for $\Gamma$ model ($\vartheta/\pi = 0.5$).

The entire quantum phase diagram of Eq.~\eqref{BMJG-Ham} is presented in Fig.~\ref{FIG-MSFTht}b.
In addition to the conventional zigzag and stripy orderings,
there is a disordered phase, which is later interpreted as a QSL,
is stabilized in a large region between $\vartheta_{t,l}$ and $\vartheta_{t,r}$
with $\vartheta_{t,l}/\pi \simeq 0.50$ and $\vartheta_{t,r}/\pi = 0.66(1)$.
It should be noticed that even though the bond-modulated Heisenberg interaction has a strong tendency to favor the zigzag ordering,
the ground state of $\Gamma$ model remains disordered albeit the transition point $\vartheta_{t,l}$ is so close to $\pi/2$.
A more elaborative study on hexagonal clusters of $N$ = 24 and 32 suggests that $\vartheta_{t,l}/\pi = 0.498(1)$
(see \underline{Supplementary Fig.~S12}).
For other less sensitive perturbations such as the third-nearest-neighbor Heisenberg ($J_3$) interaction,
the zigzag order could only exist for $J_3/\Gamma > 0.075$ (see \underline{Supplementary Fig.~S20}).
While for another off-diagonal $\Gamma'$ interaction, the zigzag order is generated at $\Gamma'/\Gamma < -0.015$\cite{LuoCSarXiv2020}.
These, in turn, confirm the robust nonmagnetic character of $\Gamma$ model notwithstanding the aggressive zigzag ordering.

\begin{figure}[!ht]
\centering
\includegraphics[width=1.00\linewidth, clip]{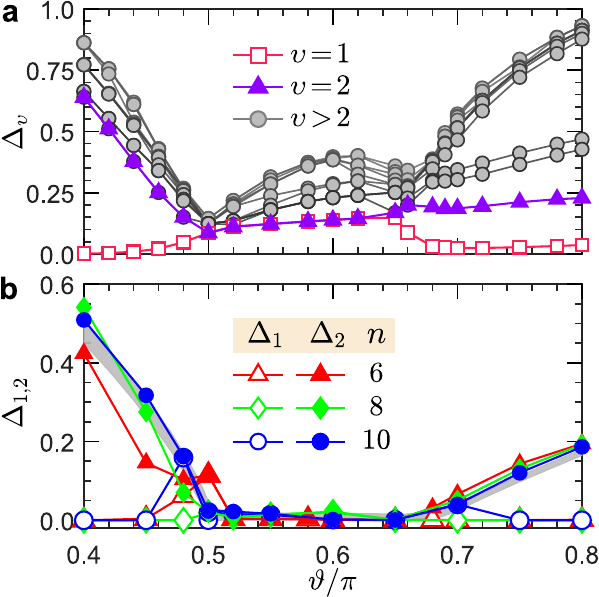}\\
\caption{\textbf{Excitation gaps.}
    \textsf{\textbf{a}} The first fifteen excitation gaps $\Delta_{\upsilon}$ with degeneracy for a $N = 24$ hexagonal cluster.
    The zigzag and stripy phases are doubly degenerate while the ground state of the QSL in the middle is unique.
    \textsf{\textbf{b}} The two lowest excitation gaps $\Delta_1$~(open symbols) and $\Delta_2$~(filled symbols) at three XC clusters where the circumferences $n$ are 6~(red triangle), 8~(green diamond), and 10~(blue circle). The thick belt is the extrapolated bulk gap.}\label{FIG-EgGAPTht}
\end{figure}

\subsection*{Gapless excitation and entropy scaling\\}
Next, we investigate the nature of the disordered phase by calculating the excitation gap and entanglement entropy.
For this purpose, we target the first sixteen energy states on a 24-site hexagonal cluster by the DMRG method (see ``Methods'' section),
and the first fifteen low-lying excitation gaps, $\Delta_{\upsilon} = E_{\upsilon}-E_g$ ($\upsilon = 1-15$), is shown in Fig.~\ref{FIG-EgGAPTht}a.
It can be seen that $\Delta_{1}$ is vanishing small while $\Delta_{2}$ survives in the zigzag/stripy phases,
indicative of the doubly degenerate ground states predicted by the semi-classical analysis.
In the intermediate region, the ground state is unique
and the density of state in the low-energy spectrum is higher than its neighbors.
Such a collapse of excitation gaps could be interpreted as a sign of gapless spectrum\cite{ZhuKSF2018,HickeyTrebst2019}.
To check the behavior of the lowest excitation gap as $N$ is varied,
we focus on four selected points at $\vartheta/\pi$ = 0.40, 0.50, 0.60, and 0.80, under hexagonal cluster of $N$ = 18, 24, and 32.
As can be seen from Tab.~\ref{Tab-HexGapN32},
the lowest excitation gap $\Delta_2$ of the zigzag phase ($\vartheta/\pi = 0.40$) and stripy phase ($\vartheta/\pi = 0.80$)
is considerably large and slightly grows with the increasing of system size.
For the intermediate phase, however, the lowest excitation gap $\Delta_1$ at $\vartheta/\pi$ = 0.50 and 0.60 declines quickly when $N$ changes from 18 to 32,
indicating that the excitation gap tends to close eventually.

\begin{table}[th!]
\caption{\label{Tab-HexGapN32}
    \textbf{Lowest excitation gap on hexagonal clusters.}
    The lowest excitation gap at $\vartheta = 0.40\pi$, $0.50\pi$, $0.60\pi$, and $0.80\pi$ on hexagon clusters of $N$ = 18, 24, and 32.
    For the zigzag/stripy phase the lowest excitation gap is $\Delta_2$, while for the intermediate region the lowest excitation gap is $\Delta_1$.}
\begin{ruledtabular}
\begin{tabular}{c c c c c}
\multicolumn{1}{c}{\multirow{2}{*}{$\vartheta/\pi$ }} &
\multicolumn{1}{c}{\multirow{2}{*}{Phases}} &
\multicolumn{3}{c}{Lowest excitation gap} \\
\cline{3-5}
\multicolumn{2}{c}{}                    & $N = 18$          & $N = 24$          & $N = 32$              \\
\hline
$0.40$              & zigzag            & --                & 0.63943681        & 0.73353505            \\
$0.50$              & QSL               & 0.08821080        & 0.08656249        & 0.04009844            \\
$0.60$              & QSL               & 0.19543027        & 0.13898662        & 0.08599835            \\
$0.80$              & stripy            & --                & 0.22834260        & 0.24451198            \\
\end{tabular}
\end{ruledtabular}
\end{table}

\begin{figure}[!ht]
\centering
\includegraphics[width=0.95\linewidth, clip]{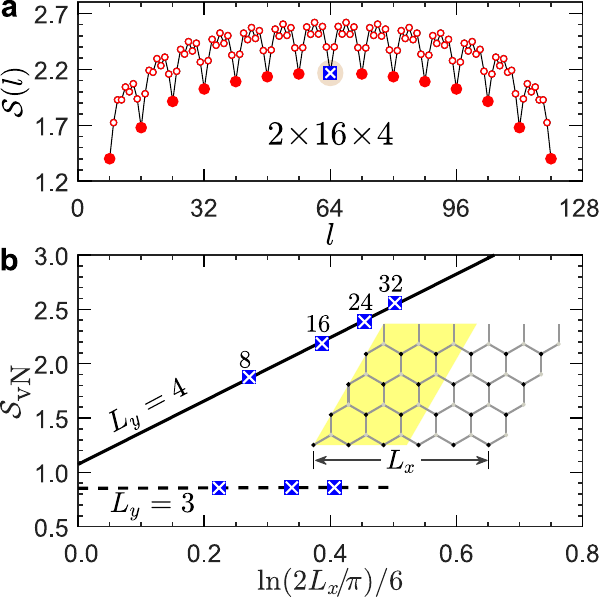}\\
\caption{\textbf{Entanglement entropy scaling.}
    \textsf{\textbf{a}} Entanglement entropy $\mathcal{S}(l)$ of a consecutive segment of length $l$ on a $2\times16\times4$ cylinder in the $\Gamma$ model.
    The solid symbols of the lowest branch represent the neat edge-cutting with $l$ being a multiply of 8 (i.e., the number of the sites along each column).
    The bipartite entanglement entropy with $l = N/2$ is marked as a blue square.
    \textsf{\textbf{b}} Entanglement entropy scaling on three-leg cylinders (dashed line) and four-leg cylinders (solid line).
    The fitting constants are $(c, c') \approx (0.02, 0.85)$ and $(c, c') \approx (2.92, 1.09)$ for $L_y$ = 3 and 4, respectively.
    The inset shows the bipartite partition of a four-leg cylinder with equal sites in the left and right halves.
    }\label{FIG-EES}
\end{figure}

We then turn to XC cylinders which enable us to calculate the excitation gaps on large system sizes.
Above all, we calculate the first fifteen excitation gaps on a XC cluster of $8\times4$
and we find that there is unlikely a big ground-state degeneracy in the intermediate region
since the gap increases gradually without abrupt change (see \underline{Supplementary Fig.~S8}).
Therefore, we only present the two lowest excitation gaps on three larger XC clusters up to 200 sites (see Fig.~\ref{FIG-EgGAPTht}b).
The gaps in the middle are rather small, indicative of a gapless region.
We also take a closer look of the gaps at $\vartheta/\pi = 0.50$ where different YC clusters are also adopted.
For either XC or YC cluster, $\Delta_{1,2}$ show a decreasing trend with the increasing of circumference $n$.
In spite of the oscillation in value, they appear to vanish within a reasonable round-off,
showing the gaplessness of the excitation spectrum in $\Gamma$ model (see \underline{Supplementary Fig.~S13}).
As a final consistency check for the gapless nature,
we calculate the excitation gap on cylinder geometry of $2\times L_x\times L_y$
(for geometry, see inset of Fig.~\ref{FIG-EES}b) with $N = 2L_xL_y$ sites in total.
Although the three-leg cylinder ($L_y = 3$) is gapped, excitation gap on cylinder of $L_y = 4$ decreases quickly with $L_x$
and is expected to disappear as $L_x\to\infty$ (see \underline{Supplementary Fig.~S14}).
Such a strong size-dependent behavior of the excitation gap is typical of gapless systems.

The entanglement entropy has appeared as a versatile tool in diagnosing quantum critical systems described by conformal field theory.
In this regard, the von Neumann entanglement entropy is introduced and it is defined as $\mathcal{S}(l) = -\textrm{Tr}(\rho\ln\rho)$
where $\rho$ is the reduced density matrix of a subregion with length $l$\cite{Eisert2010}.
Figure~\ref{FIG-EES}a shows the representative behavior of $\mathcal{S}(l)$ in the $\Gamma$ model
on a $2\times16\times4$ cylinder, which contains eight sites along each column.
When $l$ is a multiply of 8, it corresponds to a neat edge-cutting where the two halves have smooth margins.
The entanglement entropy is minimized and forms a lower branch as marked by solid symbols.
Otherwise the subsystems will be more entangled, gaining extra entropy over the lower bound.
Therefore, we shall fix $l = N/2$ to extract the central charge upon a series of finite cylinders.
For such a critical system, it is recognized that the entanglement entropy scaling takes the form of
$\mathcal{S}_{\textrm{vN}} \equiv \mathcal{S}(N/2) = \frac{c}{6}\ln\big(\frac{2L_x}{\pi}\big) + c'$
where $c$ is the central charge and $c'$ is a non-universal constant\cite{Eisert2010}.
In Fig.~\ref{FIG-EES}b, we shows the logarithmic fitting of $\mathcal{S}_{\textrm{vN}}$ for cylinders of length $L_x$ = 8, 16, 24, and 32.
It is found that $\mathcal{S}_{\textrm{vN}}$ obeys the formula well
with the fitting constants $(c, c') \approx (2.92, 1.09)$, showing that the central charge is close to 3.
An alternative fitting of the lowest branch of the entropy on each cylinder also demonstrates that $c \simeq 3$ (see \underline{Supplementary Fig.~S16}).
By contrast, for the three-leg cylinder the entropy is extremely insensitive to the length (see Fig.~\ref{FIG-EES}b),
revealing a central charge of 0.
The fact that the central charge depends highly on the width ($L_y$) of cylinders may imply the existence of spinon Fermi surface (SFS).
In this scenario, the pockets of SFS might be detected by different cuts in the Brillouin zone,
and thus the central charge could vary for different $L_y$.
We note in passing that the central charge argument has also been used to explore the possible SFS in the field-induced gapless QSL in the Kitaev model \cite{JiangWHetalarXiv2018,PatelTrivedi2019,ZouHe2020}.
Another possibility is a Dirac QSL \cite{IqbalBSP2013,HuZEH2019} with three Dirac Fermions around $\textbf{M}$ points,
which is potentially consistent with the central charge $3$ on four-leg cylinders.
Information of the central charge on wider cylinders should be helpful to distinguish between the two scenarios.
Regardless of different QSL natures, the vanishing magnetization and excitation gap in the $\Gamma$ model,
together with the distinct central charges on cylinders of $L_y = 3$ ($c = 0$) and $L_y = 4$ ($c = 3$),
manifest that its ground state is likely a gapless QSL.

\subsection*{Flux-like density and plaquette correlation\\}
So far, we have confirmed that there is no magnetic ordering in the honeycomb $\Gamma$ model,
yet little is known about the lattice symmetry breaking.
Very recently, there is a proposal of plaquette ordering stemming from a broken translational symmetry
in the \textit{classical} $\Gamma$ model\cite{SahaFZetal2019}.
it is thus of interest to examine whether there is a plaquette ordering in the quantum situation.
To this end, we study the hexagonal plaquette operator $\hat{W}_p$ and its correlation.
Actually, $\hat{W}_p$ also has its own merit as it can capture the associated phase transitions\cite{GordonCSetal2019}.
The six-body plaquette operator is known as\cite{Kitaev2006}
\begin{equation}\label{HexWp}
\hat{W}_p = 2^6 \prod_{i\in p} S_i^{\gamma} = 2^6 S_1^{x}S_2^{y}S_3^{z}S_4^{x}S_5^{y}S_6^{z},
\end{equation}
which is the product of spin operators on out-going bonds around a plaquette~(see Fig.~\ref{FIG-XC6}).

\begin{figure}[!ht]
\centering
\includegraphics[width=0.95\columnwidth, clip]{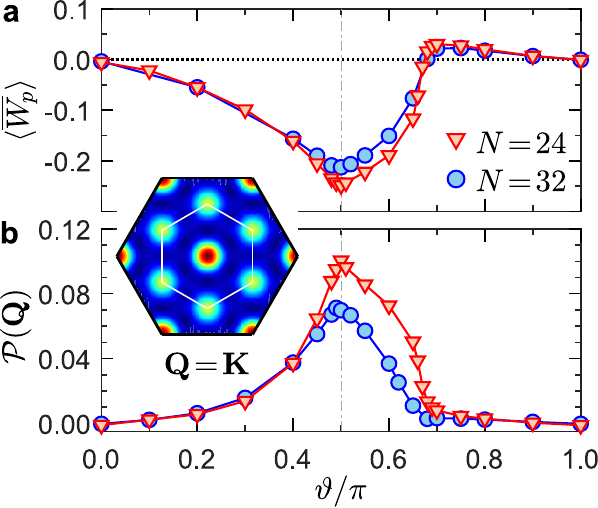}\\
\caption{\textbf{Flux-like density and plaquette correlation.}
    \textsf{\textbf{a}} Flux-like density $\langle \overline{W}_p\rangle$ and
    \textsf{\textbf{b}} plaquette order parameter $\mathcal{P}(\textbf{Q})$ for $N$ = 24 (red triangular) and 32 (blue circle).
    The inset exhibits the plaquette structure factor of $\Gamma$ model,
    which has a relatively weak peak at $\mathbf{K}$ point (corner of the Brillouin zone) in the reciprocal space.}\label{FIG-WpWp}
\end{figure}

Figure~\ref{FIG-WpWp}a shows the flux-like density $\langle \overline{W}_p\rangle = \sum_p \langle \hat{W}_p\rangle/N_p$
where $N_p = N/2$ is the number of hexagonal plaquette on clusters of $N$ = 24 and 32.
Starting from $\vartheta/\pi$ = 0.0, $\langle \overline{W}_p\rangle$ is zero,
followed by a continuous decrease before arriving at the transition point, $\vartheta/\pi \simeq 0.50$.
Afterwards, it begins to increase and then surpasses the critical line
to enter into the stripy phase where $\langle \overline{W}_p\rangle > 0$.
Recalling the quantum phase diagram shown in Fig.~\ref{FIG-MSFTht}b,
our result corroborates that the flux-like density could signal phase transitions.
For $\Gamma$ model we have $\langle \overline{W}_p\rangle = -0.25(2)$,
which is about a quarter of that in the Kitaev model\cite{Kitaev2006}.
Based on the plaquette-plaquette correlation $\langle \hat{W}_p\hat{W}_q\rangle$,
we then introduce the plaquette order parameter $\mathcal{P}_{N_p}({\bf{Q}})$ via
the plaquette structure factor $\mathcal{W}_{N_p}({\bf{Q}})$\cite{SahaFZetal2019} (see ``Methods" for definition).
In Fig.~\ref{FIG-WpWp}b, we show that the QSL phase has a vigorous peak at $\boldsymbol{\Gamma}$ point in the reciprocal space
but a weaker intensity at $\mathbf{K}$ point, signifying a perceptible plaquette correlation.
Nevertheless, The fact that the strength of $\mathcal{P}_{N_p}({\bf{K}})$ goes down rapidly suggests that
there is unlikely a plaquette ordering in the region, further corroborating a QSL without a broken symmetry.

\section*{DISCUSSION}

Ever since the seminal proposal of the Kitaev interaction in heavy $4d/5d$ transition metal oxides\cite{Jackeli2009},
which triggers the thriving research direction of Kitaev materials,
tremendous efforts have been devoted to realizing the Kitaev QSL in real materials,
yet hampered by the ineluctable non-Kitaev terms such as the off-diagonal $\Gamma$ interaction.
Whereas the honeycomb $\Gamma$ model has drawn enormous attention, its quantum nature is still under debate.
To this end, we introduce a bond-modulated Heisenberg interaction to check its tendency towards probable magnetic orderings.
For the magnetic phase diagram of the proposed bond-modulated $\tilde{J}$-$\Gamma$ model,
we find an intermediate region which is intervened between the zigzag and stripy phases.
Though exhibiting magnetic order at the classical level,
quantum fluctuations suppress such ordering since it acquires a large energy according to the spin-wave result.
In the quantum case, it turns out to be disordered and is separated from its two neighbors by first-order transitions.
By taking massive numerical efforts on the $\Gamma$ model, we are able to confirm the following three subtle physical issues.
(i) The low-energy spectrum is rather dense on a 24-site hexagonal cluster,
and the lowest excitation gap goes down gradually with the expansion of cluster size.
The empirical extrapolation on large cylinders up to 200 sites gives a vanishing energy gap,
in line with the logarithmic behaviors of entanglement entropy.
(ii) The zigzag magnetic ordering vanishes eventually,
consistent with the suppression of magnetization of the zigzag order by spin-wave analysis.
(iii) In the plaquette structure factor,
there is a perceptible short-range plaquette correlation because of a subleading peak at $\textbf{K}$ point.
These findings strongly corroborate the ground state of $\Gamma$ model
is a gapless QSL rather than a zigzag order, despite the latter being close in energy.

We would like to mention that, due to the gapless nature and for the lack of continuous spin symmetry,
it is exceedingly challenging to capture the fractionalized excitation in the proposed QSL.
The flux insertion method, which pumps fractional particles from one edge to the other,
is usually a promising way to elucidate the topological characters of the ground state.
It is performed by adiabatically twisting boundary conditions of the Hamiltonian
so that the $U(1)$ symmetry is required.
The discrete symmetries of the $\Gamma$ model thus inherently hinder this trick.
Actually, topological degeneracy is in general not well-defined for a gapless QSL,
as different gauge sectors are closely connected due to gapless excitations.
Nonetheless, the Kitaev QSL is special because flux $\hat{W}_p$ is a conserved quantity,
allowing for the identification of different flux sectors by the vison insertion\cite{HickeyTrebst2019}.
Also of note is that a recent study suggests the existence of a nematicity
due to the lattice rotational symmetry breaking\cite{GohlkeCKK2020}.
We would like to stress that the symmetry of $\Gamma$ model itself is discrete
and the asymmetrical boundary condition could cause instability on the landscape of bond energy,
making it hard to determine the nematicity in the thermodynamic limit.
However, the pending lattice nematicity does not alter our proposal of the QSL,
because it could be accompanied by a broken lattice symmetry as reported in other theoretical models\cite{GongZS2017,HuGLetal2019}.
Despite such challenges,
our work emphasizes on the inspiring and intractable quantum nature of the $\Gamma$ model.
Notably, the dominating $\Gamma$ region could be realized in $\alpha$-RuCl$_3$ under compression
where the magnitude of Kitaev interaction is small\cite{LuoCSarXiv2020}.
In short, our results provide a significant guidance to further theoretical and experimental studies on honeycomb magnets.


\section*{METHODS}

\subsection*{Density matrix renormalization group\\}
In order to check for finite-size effects,
we have performed large-scale DMRG calculations\cite{White_1992,StoudenmireWhite_2012} on three kinds of cluster geometries.
Firstly, the frequently used geometry is a $L_x \times L_y$ XC$n$ cluster under cylindrical boundary condition (see Fig.~\ref{FIG-XC6}).
Here, X indicates the orientation of the cylinder, while $n$ is the circumference of the cylinder.
We consider even circumferences $n$~($ = L_y/a_0$) ranging from 4 to 10 lattice spacing $a_0$, 
and use fixed ratio $L_x/L_y = 2$ unless stated explicitly otherwise.
$N = L_xL_y$ is the total number of spins.
Secondly, we consider the honeycomb cylinder of $2\times L_x \times L_y$
where $L_x$ ($L_y$) is the number of unit cell along $\mathbf{e}_1 = (\sqrt3, 0)$ ($\mathbf{e}_2 = (1/2, \sqrt3/2)$) direction
(see the inset of Fig.~\ref{FIG-EES}).
Due the limitation of modern computational capability, we focus primarily on four-leg cylinders ($L_y = 4$).
The maximal value of $L_x$ is 32, and the total number of spins $N = 2L_xL_y$.
Lastly, we also use the $C_3$ symmetric hexagonal cluster with $N$ = 24 or 32 sites under full periodic boundary conditions.
In all cases, we keep up to $m$ = 3000$\sim$5000 states and perform about 12 sweeps in the calculation
so as to ensure the truncation error is smaller than $10^{-6}$.
When targeting the first few low-lying energy levels,
we diagonalize a subspace of a sparse Hermitian matrix iteratively by Davidson algorithm,
with the precision of each eigenvalue maintained at a desired standard.
In addition, all the targeted states are used with an equal weight to construct the reduced density matrix.

The magnetic order parameter is defined by $M_N({\bf{Q}}) = \sqrt{\mathbb{S}_N({\bf{Q}})/N}$
where $\mathbb{S}_N({\bf{Q}}) = \sum_{\alpha\beta}\delta_{\alpha\beta}\mathbb{S}_N^{\alpha\beta}({\bf{Q}})$
is the total static magnetic structure factor, with
\begin{equation}
\mathbb{S}_N^{\alpha\beta}({\bf{Q}})=\frac{1}{N}\sum_{ij}
    \langle{S^{\alpha}_i {S^{\beta}_j}}\rangle e^{i{\bf{Q}}\cdot{({\bm{R}}_i-{\bm{R}}_j)}}.
\end{equation}
Here, ${\bm{R}}_i$ is the position of spin and ${\bf{Q}}$ is the ordering wavevector.
We also calculate the plaquette-plaquette correlation $\langle \hat{W}_p \hat{W}_q\rangle$
where $\hat{W}_p$ is the hexagon plaquette operator (see Eq.~\eqref{HexWp}).
Likewise, we define the static plaquette structure factor
\begin{equation}\label{EQSM-VisonSF}
\mathcal{W}_{N_p}({\bf{Q}})=\frac{1}{N_p}\sum_{pq}
\langle{\hat{W}_p \hat{W}_q}\rangle e^{i{\bf{Q}}\cdot{({\bm{R}}_p-{\bm{R}}_q)}},
\end{equation}
where $\bm{R}_p$ is the central position of each plaquette,
and $N_p = N/2$ is the number of plaquette.
To eliminate the strong finite-size effect due to the identity $\big\langle (\hat{W}_p)^2\big\rangle$ = 1,
we introduce the plaquette order parameter (see \underline{Supplementary Note 5})
\begin{equation}\label{EQ-HiddenOPPlqtt}
\mathcal{P}_{N_p} = \sqrt{\frac{\mathcal{W}_{N_p}({\bf{Q}})}{N_p}} - \frac{1}{\sqrt{N_p}}.
\end{equation}

\subsection*{Simulation and energy optimization\\}
We use the parallel tempering Monte Carlo simulation with the heat-bath algorithm to prevent the possible metastable state
at low temperatures\cite{Metropolis1953,HukushimaNemoto1996}.
The simulation is carried out in a temperature range with a hundred of replicas.
The heat-bath algorithm is performed at given temperature,
followed by a so-called thermal replicas where configuration swaps between different temperatures are allowed
with a probability according to a detailed balance condition.
The simulations are performed on three XC clusters of $16\times16$, $24\times24$, and $32\times32$, under toroidal boundary condition.

The resulting energy-optimized spin configuration is then served as the benchmark for the analytical calculation.
The classical spin can be written as
\begin{equation}\label{Mclassicalspin3}
\boldsymbol{S}_i = S\left(\sin{\theta_i}\cos{\phi_i}, \sin{\theta_i}\sin{\phi_i}, \cos{\theta_i}\right),
\end{equation}
where $\theta_i \in [0, \pi)$ and $\phi_i \in [0, 2\pi)$.
Taking the zigzag (zz) order and stripy (st) order for instance, their classical energy are
\begin{equation}\label{MCEg-ZZ}
e_g^{\textrm{zz}} = -\frac12\big(3\tilde{J} + \mathcal{F}_{\max}(\theta, \phi)\Gamma \big)
\end{equation}
and
\begin{equation}\label{MCEg-ST}
e_g^{\textrm{st}} = \frac12\big(3\tilde{J} + \mathcal{F}_{\min}(\theta, \phi)\Gamma \big)
\end{equation}
where the explicit form of the auxiliary function is
\begin{equation}\label{MCEg-FuncThtPhi}
\mathcal{F}(\theta, \phi) = \sin^2\theta\sin2\phi - \sin2\theta(\sin\phi + \cos\phi).
\end{equation}
Mathematically, the maximum of Eq.~\eqref{MCEg-FuncThtPhi} is 2
with $(\theta, \phi) = (\pi-\mathrm{atan}(\sqrt2), \pi/4)$ or $(\theta, \phi) = (\mathrm{atan}(\sqrt2), 5\pi/4)$.
This means that the classical energy of the zigzag phase is $e_g^{\textrm{zz}} =  -(2\Gamma+3\tilde{J})/2$
with the classical magnetic direction $\mathbf{n} = [11\bar1]$.
The mimimum of Eq.~\eqref{MCEg-FuncThtPhi} is $-1$, and the energy of the stripy phase is $e_g^{\textrm{st}} =  -(\Gamma-3\tilde{J})/2$.
Its moment direction is free to vary in a plane that is perpendicular to $\mathbf{n}$,
indicating an emergent continuous symmetry in the classical stripy phase.
The analytical energy and spin configurations of the other magnetic phases are shown in \underline{Supplementary Note 1}.

\subsection*{Linear spin-wave theory\\}
We summarize the derivation of the spin wave spectra for the zigzag phase,
which is one of the degenerate ground states of the classical $\Gamma$ model.
In the framework of linear spin-wave theory,
the local spin operator $\mathbf S_i = \left(S_i^x, S_i^y, S_i^z\right)$ is represented by
bosonic creation and annihilation operators $a_i$ and $a_i^{\dagger}$.
By virtue of the Holstein-Primakoff transformation, we have
\begin{eqnarray}\label{HPTran}
	\tilde{S}_i^+ \simeq \sqrt{2S} a_i,\quad
	\tilde{S}_i^- \simeq \sqrt{2S} a_i^\dagger,\quad
	\tilde{S}_i^n = S - a_i^\dagger a_i.
\end{eqnarray}
Here, $\tilde{S}_i^n \equiv (\mathbf S \cdot \mathbf n)$ is the spin component along the classical ordered moment $\mathbf n$
and $\tilde{S}_i^\pm \equiv (\mathbf S_i \cdot \mathbf e) \pm \imath [\mathbf S_i \cdot (\mathbf n \times \mathbf e)]$
are the ladder operators consisting of the orthogonal spin components,
with $\mathbf e$ being an (arbitrary) unit vector perpendicular to $\mathbf n$ and satisfying the right-hand rule\cite{JanssenVojta2019}.
The spin operator is thus
\begin{eqnarray}\label{SpinOpN_E_NxE}
\mathbf{S}_{\tau,i}
&=& \sqrt{\frac{S}{2}}(a_i+a_i^{\dagger}){\mathbf{e}} + \tau\sqrt{\frac{S}{2}}(-\ii a_i+\ii a_i^{\dagger}) (\mathbf n \times \mathbf e) \nonumber\\
& & +\tau(S-a_i^{\dagger}a_i) {\mathbf{n}}
\end{eqnarray}
where $\tau$ is introduced for classical spin which is either parallel ($\tau = +1$) or antiparallel ($\tau = -1$) to $\mathbf n$.
The $\gamma$-component of the spin $S_{\tau,i}^{\gamma} = \mathbf{S}_{\tau,i} \cdot \mathbf{e}_{\gamma}$.

For the zigzag phase, we choose the following orthogonal axis,
$\mathbf{e} = [112]$, $\mathbf{n}\times\mathbf{e} = [1\bar10]$, and $\mathbf{n} = [11\bar1]$.
Because of the four-sublattice ($n_s$ = 4) nature,
the magnetic unit cell is taken as the rectangle of the area $a_1 \times a_2$ with $a_1 = 3 a_0$ and $a_2 = \sqrt3 a_0$,
see Fig.~\ref{FIG-XC6}a.
Within the magnetic unit cell, the wave vector $\bm q$ could be parameterized in units of $(h, k)$ as
$\bm q$ = $\big(\frac{2\pi}{a_1}h, \frac{2\pi}{a_2}k\big)$\cite{ChoiCKetal2012}.
In this notation, $\mathbf{M}$ and $\mathbf{M}'$ points in the Brillouin zone could be rewritten as (1, 0) and (0, 1), respectively.
By introducing four flavors of Holstein-Primakoff bosons and using the Fourier transformation,
we arrive at the following spin-wave Hamiltonian
\begin{eqnarray}\label{HswZZ}
\mathcal{H}_{SW}^{\mathrm{zz}} = -N S(S+1) e_g^{\textrm{zz}} + \frac{S}{2} \sum_{\bm q}
    \hat{\bf x}_{\bm q}^\dagger
    \hat{\bf H}_{\bm q}
    \hat{\bf x}_{\bm q}^{\phantom{\dagger}},
\end{eqnarray}
where $\hat{\bf x}^\dag_{\bm q}=\big( a^\dag_{\bm q}, b^\dag_{\bm q}, c^\dag_{\bm q}, d^\dag_{\bm q},
a^{\phantom \dag}_{\bm{-q}},b^{\phantom \dag}_{\bm{-q}}, c^{\phantom \dag}_{\bm{-q}}, d^{\phantom \dag}_{\bm{-q}}\big)$
is a vector of length $2 n_s$ and $\hat{\bf H}_{\bm q}$ is a $2 n_s\times 2 n_s$ matrix of the form
\begin{eqnarray}
\hat{\bf H}_{\bm q} =
\left(\begin{array}{cc}
	\hat{\Lambda}_{\bm q}              &    \hat{\Delta}_{\bm q}       \\
    \hat{\Delta}_{\bm q}^{\dagger}     &    \hat{\Lambda}_{-\bm q}^T
\end{array}\right)
\end{eqnarray}
with
\begin{eqnarray}\label{HsubDiag}
\hat{\Lambda}_{\bm q} =
\left(\begin{array}{cccc}
	A      & {E}_{\bm q}    &   \circ       &   B_{\bm q}               \\
    {E}^{*}_{\bm q}  &   A   &   B^{*}_{\bm q}   &   \circ          \\
    \circ   & B_{\bm q} &   A   &   {E}_{\bm q}           \\
    B^{*}_{\bm q}   & \circ &   {E}^{*}_{\bm q}  &  A          \\
\end{array}\right)
\end{eqnarray}
and
\begin{eqnarray}\label{HsubODiag}
\hat{\Delta}_{\bm q} =
\left(\begin{array}{cccc}
	\circ   &   C_{\bm q}   &     \circ       &   D_{\bm q, +}    \\
    C^*_{\bm q}   &   \circ   &   D^*_{\bm q, +}  &      \circ       \\
    \circ   &   D_{\bm q, -}  &   \circ  &  C_{\bm q}         \\
    D^*_{\bm q, -}  &   \circ  &  C^*_{\bm q}  &  \circ        \\
\end{array}\right).
\end{eqnarray}
The parameters in Eq.~\eqref{HsubDiag} and Eq.~\eqref{HsubODiag} are given by
\begin{align}
\left\{
  \begin{array}{l}
    A = 3\tilde{J} + 2\Gamma     \\
    B_{\bm q} = -2(\tilde{J}-\Gamma/3) \varrho^{-1} \cos\pi k \\
    C_{\bm q} = (\tilde{J}-\Gamma/3) \varrho^{2}  \\
    D_{\bm q, \tau} = \frac{2\Gamma}{3}(\cos\pi k + \tau\sqrt3 \sin\pi k) \varrho^{-1}   \\
    E_{\bm q} = \frac{2\Gamma}{3} \varrho^{2}
  \end{array}
\right.
\end{align}
where $\varrho = e^{\ii\pi h/3}$.
Since
$B_{-\bm q} = B^{*}_{\bm q}$,
$C_{-\bm q} = C^{*}_{\bm q}$,
$E_{-\bm q} = E^{*}_{\bm q}$, and
$D_{-\bm q, \tau} = D^{*}_{\bm q, -\tau}$,
we hence deduce that
$\hat{\Delta}_{\bm q}^{\dagger} = \hat{\Delta}_{\bm q}$
and
$\hat{\Lambda}_{-\bm q}^T = \hat{\Lambda}_{\bm q}^{\dagger}$.

The quadratic Hamiltonian Eq.~\eqref{HswZZ} can be diagonalized via a bosonic Bogoliubov transformation $T(\bm q)$.
To preserve the canonical commutation rules of the bosons, it should satisfy the orthogonality relations
$T \Sigma T^\dagger = T^\dagger \Sigma T = \Sigma$ where $\Sigma = \diag{\mathds 1, -\mathds 1}$.
The eigenvalues of $\Sigma\hat{\bf H}_{\bm q}$ give the magnon spectrum
$\Omega(\bm q) = \diag{\omega_{\bm q, 1}, \omega_{\bm q, 2}, \cdots, \omega_{{\bm q}, n_s}}$.
The spin-wave dispersions of other magnetic orderings (including the mixed phase and the noncollinear phase)
are shown in \underline{Supplementary Note 2}.

\section*{DATA AVAILABILITY}
\noindent The data that support the findings of this study are available from the corresponding authors upon reasonable request.


\section*{ACKNOWLEDGEMENTS}
\noindent We thank M. Gohlke, J.~S. Gordon, Y.-Z. Huang, T. Li, H.-J. Liao, K. Liu, Z.-X. Liu, B. Xi, W. Yu,
and especially S. Hu for many inspiring and helpful discussions.
X.W. was supported by the National Program on Key Research Project (Grant No. 2016YFA0300501)
and by the National Natural Science Foundation of China (Grant No. 11974244).
J.Z. was supported by the the National Natural Science Foundation of China (Grant No. 11874188).
H.-Y.K. was supported by the NSERC Discovery Grant No. 06089-2016 and acknowledged funding from the Canada Research Chairs Program.
X.W. also acknowledged additional support from a Shanghai talent program.
The computations were mostly performed on the Tianhe-2JK at the Beijing Computational Science Research Center (CSRC).

\section*{COMPETING INTERESTS}
\noindent The authors declare no competing interests.

\section*{AUTHOR CONTRIBUTIONS}
\noindent J.Z. and X.W. initiated and supervised the project.
Q.L. performed the spin-wave analysis and numerical calculations.
H.-Y.K. guided the analysis of entanglement entropy scaling and plaquette correlation.
J.Z., H.-Y.K., and X.W. checked the calculations.
All authors together discussed the numerical details and drafted the article.

\section*{ADDITIONAL INFORMATION}
\noindent \textbf{Supplementary information} is available in the online version of the paper.\\
\noindent \textbf{Correspondence} and requests for materials should be addressed to J.Z. or X.W.


%




\clearpage

\onecolumngrid


\newpage
\setcounter{page}{1}

\setcounter{equation}{0}
\setcounter{figure}{0}
\setcounter{table}{0}
\setcounter{subsection}{0}

\renewcommand{\theequation}{S\arabic{equation}}
\renewcommand{\thefigure}{S\arabic{figure}}
\renewcommand{\thetable}{S\arabic{table}}
\renewcommand{\bibnumfmt}[1]{[#1]}
\renewcommand{\citenumfont}[1]{#1}

\let\subsection\oldsection         

\renewcommand{\thesubsection}{\arabic{subsection}}


\begin{center}
{\large{\bf Supplemental Information for\\
 ``Gapless quantum spin liquid in a honeycomb $\Gamma$ magnet''}}
\end{center}
\begin{center}

Qiang Luo$^{1, 2}$, Jize Zhao$^{3}$, Hae-Young Kee$^{2, 4}$, and Xiaoqun Wang$^{5, 6}$\\
\quad\\
$^1$\textit{Department of Physics, Renmin University of China, Beijing 100872, China}\\
$^2$\textit{Department of Physics, University of Toronto, Toronto, Ontario M5S 1A7, Canada}\\
$^3$\textit{School of Physical Science and Technology $\&$ Key Laboratory for Magnetism and\\
Magnetic Materials of the MoE, Lanzhou University, Lanzhou 730000, China}\\
$^4$\textit{Canadian Institute for Advanced Research, Toronto, Ontario, M5G 1Z8, Canada}\\
$^5$\textit{Key Laboratory of Artificial Structures and Quantum Control (Ministry of Education),\\ Shenyang National Laboratory for Materials Science,\\ School of Physics and Astronomy, Tsung-Dao Lee Institute, Shanghai Jiao Tong University, Shanghai 200240, China}\\
$^6$\textit{Beijing Computational Science Research Center, Beijing 100084, China}
\end{center}





\vspace{0.50cm}
\begin{center}
\textsf{\textbf{Supplementary Note 1: Classical energy and spin configurations}}
\end{center}
\setcounter{subsection}{0}

\subsection{The mixed antiferromagnetic--twining zigzag phase}
\begin{figure}[!ht]
\centering
\includegraphics[width=0.55\columnwidth, clip]{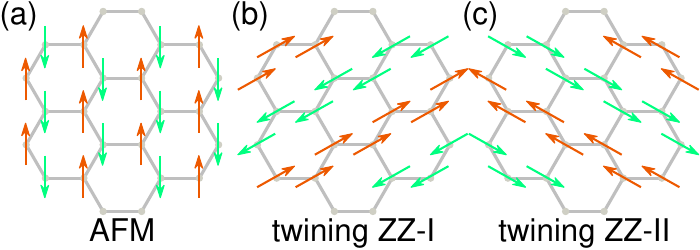}\\
\renewcommand{\figurename}{\textbf{Supplementary Figure}}
\caption{The degenerate ground states of the mixed phase, which includes
(a)~antiferromagnetic (AFM) order and (b)/(c) two kinds of twining zigzag orders.}\label{FIGSM-Mixed}
\end{figure}

Due to the bond-modulated $\eta_{\gamma} (\pm1)$ term, the $\tilde{J}$-$\Gamma$ model shown in
Eq.~(\textcolor{red}{1}) in the main text does not posses $C_6$ rotational symmetry.
This leads to a discrepancy among the three conventional configurations of the zigzag ordering with different orientions.
We find that the other two twining zigzag orders~(see Fig.~\ref{FIGSM-Mixed}(b) and (c)) have higher energy than
the standard zigzag order (see Fig.~\textcolor{red}{2}(b) of the main text) when $\vartheta/\pi \in [0, 1/2)$.
When $\vartheta/\pi$ is slightly larger than $1/2$, the twining zigzag orders overcome the latter and become the ground state.
Interestingly, the antiferromagnetic (AFM) order~(see Fig.~\ref{FIGSM-Mixed}(a)) has the same energy and contributes to the degenerate manifolds.
Moreover, due to the two nonequivalent sites per unit cell, each configuration is two-fold degenerate.
Consequently, we conclude that the mixed phase has six-fold degenerate ground states.

\begin{figure}[!ht]
\centering
\includegraphics[width=0.45\columnwidth, clip]{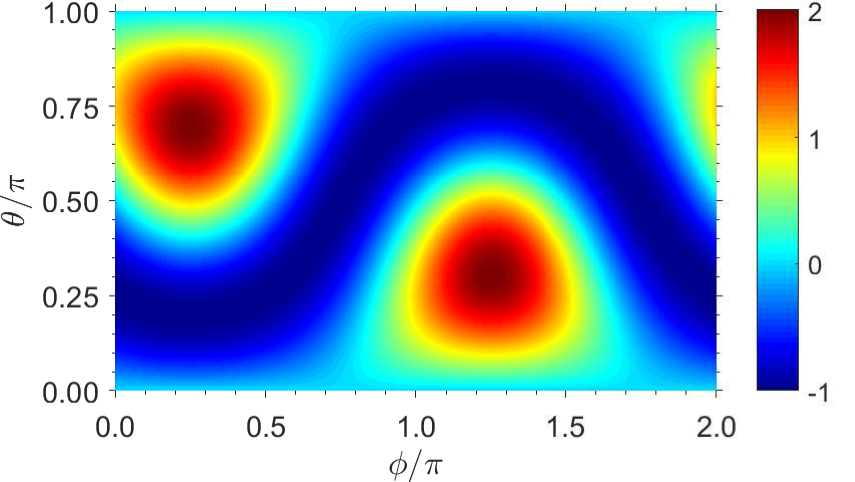}\\
\renewcommand{\figurename}{\textbf{Supplementary Figure}}
\caption{Illustration of function $\mathcal{F}(\theta, \phi)$ in the ($\theta$, $\phi$) parameter space.}\label{FIGSM-FuncForEg}
\end{figure}

For the AFM ordering out of the mixed phase, its energy is given by
\begin{equation}\label{CEg-Mixed}
e_g^{\textrm{mixed}} = -\frac12\big(-\tilde{J} + \mathcal{F}_{\max}(\theta, \phi+\pi)\Gamma \big),
\end{equation}
where $\mathcal{F}(\theta, \phi) = \sin^2\theta\sin2\phi - \sin2\theta(\sin\phi + \cos\phi)$.
The function $\mathcal{F}(\theta, \phi)$ is plotted in Fig.~\ref{FIGSM-FuncForEg},
which has a maximum of 2.
Since the translation of $\phi$ by $\pi$ does not change the magnitude of the function $\mathcal{F}(\theta, \phi)$,
we then obtain $e_g^{\textrm{mixed}} =  -(2\Gamma-\tilde{J})/2$
with $(\theta, \phi) = (\mathrm{atan}(\sqrt2), \pi/4)$
or $(\theta, \phi) = (\pi-\mathrm{atan}(\sqrt2), 5\pi/4)$.
In this case the classical magnetic direction of the AFM ordering is $\mathbf{n} = [111]$.

\subsection{The noncollinear phase}
\begin{figure}[!ht]
\centering
\includegraphics[width=0.48\columnwidth, clip]{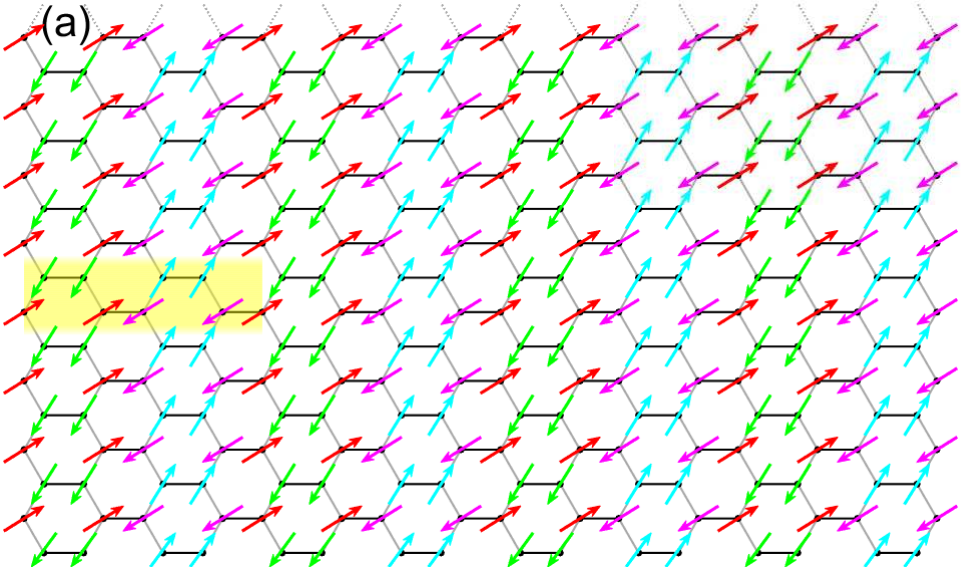}
\hspace{0.10cm}
\includegraphics[width=0.48\columnwidth, clip]{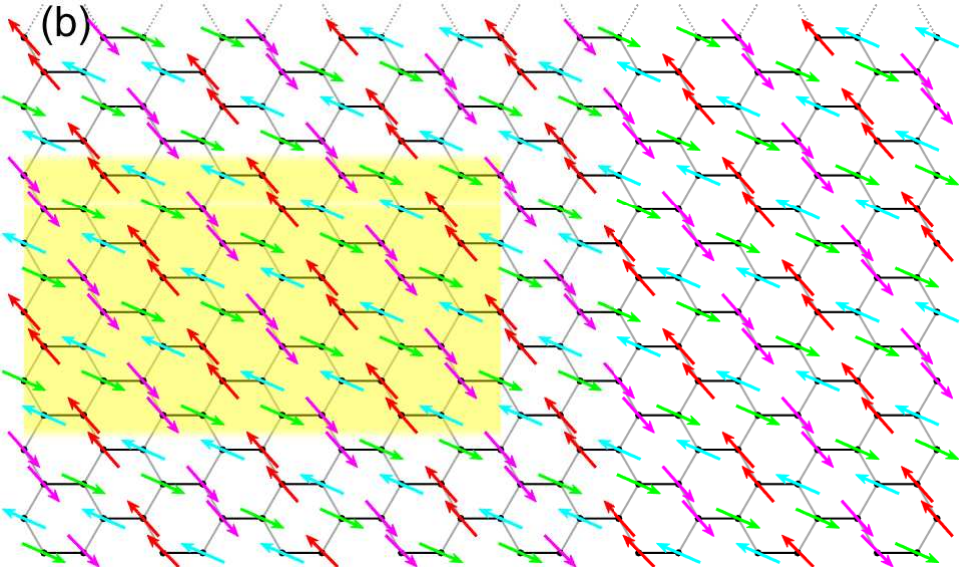}\\
\renewcommand{\figurename}{\textbf{Supplementary Figure}}
\caption{Configurations of the noncollinear phase in the real space. The size of the unit cell
    marked by the yellow shadow is (a) $4\times2$ and (b) $8\times8$ (enlarged).}\label{FIGSM-NonCop}
\end{figure}

When $\theta/\pi \approx 0.65$
there is a noncollinear~(NCL) phase consisting of two kinds of spins
(or four if we consider that two of them are anti-parallel to their partners)
which are neither (anti-)parallel nor perpendicular in the classical phase diagram.
Suppose that the angles of one kind of the spins are $(\theta, \phi)$,
then they are $(\theta+\pi/2, \phi+\pi)$ for the other. Namely,
\begin{align}
\left\{
  \begin{array}{l}
    \boldsymbol{S}_a = S\left(\sin{\theta}\cos{\phi}, \sin{\theta}\sin{\phi}, \cos{\theta}\right) \\
    \boldsymbol{S}_b =-S\left(\cos{\theta}\cos{\phi}, \cos{\theta}\sin{\phi}, \sin{\theta}\right) \\
  \end{array}
\right..
\end{align}
One of the configurations of the spins is shown in Fig.~\ref{FIGSM-NonCop}(a).
The optimal angles could be obtained by minimizing the classical energy
\begin{equation}\label{CEg-Mixed}
e_g^{\textrm{NCL}} = \tilde{J}\sin2\theta + \frac{\Gamma}{4}\big[\cos2\theta\sin2\phi - 2(\sin\phi+\cos\phi)\big].
\end{equation}
For example,
if $\theta = 3\pi/4 - \psi_0/2$ and $\phi = \pi/4$
where $\psi_0 = \mathrm{atan}\big(\frac{\Gamma}{4J}\big)$,
we have the classical energy
\begin{equation}\label{CEg-Mixed}
e_g^{\textrm{NCL}} = -\sqrt{\tilde{J}^2+\frac{\Gamma^2}{16}} - \frac{\Gamma}{\sqrt2}.
\end{equation}
We also note that the angles between the two kinds of spins are
$-\psi_0$ or its supplementary angle $\pi+\psi_0$.
Since $\psi_0$ is $\vartheta$-dependent,
the polar angle $\theta$ also varies with $\vartheta$.
The fascinating character of the noncollinear phase is that
it may also possess other ground states with larger unit cell.
For example, we find such a ground state whose (enlarged) unit cell has 64 lattice sites, see Fig.~\ref{FIGSM-NonCop}(b).

\vspace{0.50cm}
\begin{center}
\textsf{\textbf{Supplementary Note 2: Quantum fluctuations in the Linear spin-wave theory}}
\end{center}
\setcounter{subsection}{0}

Here, we go beyond the classical level by considering the quantum fluctuations to find out
where the nonmagnetic state may appear in the phase diagram.
To this end, we utilize the linear spin-wave theory
where each local spin operator $\mathbf S_i = \left(S_i^x, S_i^y, S_i^z\right)$ is represented by
bosonic creation and annihilation operators $a_i$ and $a_i^{\dagger}$.
We adopt some of the notations used by Janssen and Vojta \cite{SMJanssenVojta2019}.
By virtue of the Holstein-Primakoff transformation,
\begin{eqnarray}\label{HPTran}
	\tilde{S}_i^+ & = & \sqrt{2S} \sqrt{1 - {a_i^\dagger a_i}/{(2S)}} a_i = \sqrt{2S} a_i + \mathcal O(1/\sqrt{S}), \nonumber \\
	\tilde{S}_i^- & = & \sqrt{2S} a_i^\dagger \sqrt{1 - {a_i^\dagger a_i}/{(2S)}} = \sqrt{2S} a_i^\dagger + \mathcal O(1/\sqrt{S}), \\
	\tilde{S}_i^n & = & S - a_i^\dagger a_i. \nonumber
\end{eqnarray}
Here, $\tilde{S}_i^n \equiv (\mathbf S \cdot \mathbf n)$ is the spin component along the classical ordered moment $\mathbf n$
and $\tilde{S}_i^\pm \equiv (\mathbf S_i \cdot \mathbf e) \pm \imath [\mathbf S_i \cdot (\mathbf n \times \mathbf e)]$ are the ladder operators consisting of the orthogonal spin components,
with $\mathbf e$ being an (arbitrary) unit vector perpendicular to $\mathbf n$ and satisfying the right-hand rule.
The spin operator is thus
\begin{align}\label{SpinOpN_E_NxE}
\mathbf{S}_{\tau,i} =
& \sqrt{\frac{S}{2}}(a_i+a_i^{\dagger}){\mathbf{e}} + \tau\sqrt{\frac{S}{2}}(-\ii a_i+\ii a_i^{\dagger}) (\mathbf n \times \mathbf e)
  +\tau(S-a_i^{\dagger}a_i) {\mathbf{n}}
\end{align}
where $\tau$ is introduced for classical spin which is either parallel ($\tau = +1$) or antiparallel ($\tau = -1$) to $\mathbf n$.
For each $\gamma$-component $S_{\tau,i}^{\gamma} = {\mathbf S_i}\cdot\mathbf{e}_{\gamma}$, we have
\begin{align}\label{SpinOpABCofGamma}
S_{\tau,i}^{\gamma} &=
 \sqrt\frac{S}{2} a_i(\mathbf{e}\cdot\mathbf{e}_{\gamma} -\ii\tau(\mathbf n \times \mathbf e)\cdot\mathbf{e}_{\gamma})
 + \sqrt\frac{S}{2} a_i^{\dagger}(\mathbf{e}\cdot\mathbf{e}_{\gamma} +\ii\tau(\mathbf n \times \mathbf e)\cdot\mathbf{e}_{\gamma})
 + \tau(S-a_i^{\dagger}a_i) (\mathbf{n}\cdot\mathbf{e}_{\gamma}) \nonumber \\
&= \sqrt\frac{S}{2} a_i(\xi^{\gamma} -\ii\tau\eta^{\gamma}) + \sqrt\frac{S}{2} a_i^{\dagger}(\xi^{\gamma} +\ii\tau\eta^{\gamma}) + \tau(S-a_i^{\dagger}a_i)\zeta^{\gamma}
\end{align}
where $\xi^{\gamma} = \mathbf{e}\cdot\mathbf{e}_{\gamma}$, $\eta^{\gamma} = (\mathbf n \times \mathbf e)\cdot\mathbf{e}_{\gamma}$,
and $\zeta^{\gamma} = \mathbf{n}\cdot\mathbf{e}_{\gamma}$.
Before going into detail, we shall define an auxiliary function
\begin{equation}
\mathcal{G}_{\tau}^{\alpha\beta}(x) = (\xi^{\alpha}\xi^{\beta} + x\eta^{\alpha}\eta^{\beta}) +\ii\tau\frac{1-x}{2} (\xi^{\alpha}\eta^{\beta}+\xi^{\beta}\eta^{\alpha}),
\end{equation}
which satisfies $\mathcal{G}_{-\tau}^{\alpha\beta}(x) = \big[\mathcal{G}_{\tau}^{\alpha\beta}(x)\big]^{*}$.
Specifically,
$\mathcal{G}_{\tau}^{\alpha\beta}(1) = \mathcal{G}_{\pm}^{\alpha\beta}(1) = \xi^{\alpha}\xi^{\beta} + \eta^{\alpha}\eta^{\beta}$
and
$\mathcal{G}_{\tau=1}^{\alpha\beta}(-1) = (\xi^{\alpha} +\ii \eta^{\alpha})(\xi^{\beta} +\ii \eta^{\beta})
= \big[\mathcal{G}_{\tau=-1}^{\alpha\beta}(-1)\big]^{*}$.

\subsection{four-sublattice stripy order}
Like the zigzag order, the stripy order also has four sublattices.
Following a very similar procedure shown in ``\textsf{Method: Linear spin-wave theory}'' in the main text,
we get the spin-wave Hamiltonian for the stripy order as
\begin{eqnarray}\label{HtotAACCST}
\mathcal{H} = 2N_s S(S+1) \big[3\tilde{J} - 2\Gamma (\zeta^{y}\zeta^{z} + \zeta^{z}\zeta^{x} - \zeta^{x}\zeta^{y})\big] + \frac{S}{2} \sum_{\mathbf q}
    \hat{\bf x}_{\bf q}^\dagger
    \hat{\bf H}_{\bf q}
    \hat{\bf x}_{\bf q}^{\phantom{\dagger}},
\end{eqnarray}
where $\hat{\bf x}^\dag_{\bf q}=\left( a^\dag_{\bf q}, b^\dag_{\bf q}, c^\dag_{\bf q}, d^\dag_{\bf q},
a^{\phantom \dag}_{\bf -q},b^{\phantom \dag}_{\bf -q}, c^{\phantom \dag}_{\bf -q}, d^{\phantom \dag}_{\bf -q}\right)$
is a vector of length $2 n_s$ and $\hat{\bf H}_{\bf q}$ is a $2 n_s\times 2 n_s$ matrix of the form
\begin{eqnarray}\label{HtotZZBlk}
\hat{\bf H}_{\bf q} =
\left(\begin{array}{cc}
	\hat{\Lambda}_{\bf q}              &    \hat{\Delta}_{\bf q}       \\
    \hat{\Delta}_{\bf q}^{\dagger}     &    \hat{\Lambda}_{-\bf q}^T
\end{array}\right)
\end{eqnarray}
with
\begin{eqnarray}\label{HtotAACC}
\hat{\Lambda}_{\bf q} =
\left(\begin{array}{cccc}
	A      & {E}_{\mathbf q}    &   \circ       &   B_{\mathbf q, +}               \\
    {E}^{*}_{\mathbf q}  &   A   &   B^{*}_{\mathbf q, -}   &   \circ          \\
    \circ   & B_{\mathbf q, -} &   A   &   {E}_{\mathbf q}           \\
    B^{*}_{\mathbf q, +}   & \circ &   {E}^{*}_{\mathbf q}  &  A          \\
\end{array}\right),\quad
\hat{\Delta}_{\bf q} =
\left(\begin{array}{cccc}
	\circ   &   C_{\mathbf q, +}   &     \circ       &   D_{\mathbf q}    \\
    C_{-\mathbf q, +}   &   \circ   &   D_{-\mathbf q}  &      \circ       \\
    \circ   &   D_{\mathbf q}  &   \circ  &  C_{\mathbf q, -}         \\
    D_{-\mathbf q}  &   \circ  &  C_{-\mathbf q, -}  &  \circ        \\
\end{array}\right).
\end{eqnarray}
Those matrix entries are given by
\begin{align}\label{EQSM-STABCDEForm}
\left\{
  \begin{array}{l}
    A = 2\Gamma(\zeta^{y}\zeta^{z} + \zeta^{z}\zeta^{x} - \zeta^{x}\zeta^{y}) - 3\tilde{J} \\
    B_{\mathbf q, \tau} = \Gamma \big(\mathcal{G}_{\tau}^{yz}(-1)e^{i{\mathbf q}\boldsymbol\delta_{x}} + \mathcal{G}_{\tau}^{zx}(-1)e^{i{\mathbf q}\boldsymbol\delta_{y}}\big) \\
    C_{\mathbf q, \tau} = \Gamma \mathcal{G}_{\tau}^{xy}(-1) e^{i{\mathbf q}\boldsymbol\delta_{z}}  \\
    D_{\mathbf q} = -\tilde{J} (e^{i{\mathbf q}\boldsymbol\delta_{x}} + e^{i{\mathbf q}\boldsymbol\delta_{y}})
        + \Gamma \big(\mathcal{G}^{yz}(1)e^{i{\mathbf q}\boldsymbol\delta_{x}}
        + \mathcal{G}^{zx}(1)e^{i{\mathbf q}\boldsymbol\delta_{y}}\big)   \\
    E_{\mathbf q} = \big[\tilde{J} + \Gamma \mathcal{G}^{xy}(1)\big] e^{i{\mathbf q}\boldsymbol\delta_{z}}
  \end{array}
\right.,
\end{align}
where $\boldsymbol\delta_{x} = \big(-\frac{a_1}{6}, \frac{a_2}{2}\big)$,
$\boldsymbol\delta_{y} = \big(-\frac{a_1}{6}, -\frac{a_2}{2}\big)$,
and $\boldsymbol\delta_{z} = \big(\frac{a_1}{3}, 0\big)$.
Here, $a_1 \times a_2$ is the unit cell of the zigzag/stripy order with $a_1 = 3$ and $a_2 = \sqrt3$,
see Fig.~\textcolor{red}{1} of the main text.
For simplicity, the wave vector $\bm q$ is parameterized in units of $(h, k)$ as
$\bm q$ = $\big(\frac{2\pi}{a_1}h, \frac{2\pi}{a_2}k\big)$\cite{SMChoiCKetal2012}.

Actually, there is an emergent continuous $U(1)$ symmetry for the stripy order.
The spins are perpendicular to $\mathbf{\tilde c}~[11\bar{1}]$,
but could vary freely in the plane spanned by $\mathbf{\tilde a}~[112]$ and $\mathbf{\tilde b}~[1\bar{1}0]$.
As demonstrated in Fig.~\textcolor{red}{3}(b) in the main text, these degeneracy is lifted by quantum fluctuations,
leading to a two-fold degenerate ground state which is either parallel or antiparallel to $\mathbf{\tilde b}$.
Thus, we choose $\mathbf{n} = \mathbf{\tilde b}$ as the magnetically ordered moment direction
and $\mathbf{e} = \mathbf{\tilde c}$ as the arbitrary unit vector,
then we find that the explicit formula in Eq.~\eqref{EQSM-STABCDEForm} is given by
\begin{align}\label{EQSM-STABCDERslt}
\left\{
  \begin{array}{l}
    A = \Gamma - 3\tilde{J}     \\
    B_{\mathbf q, \tau} = \frac{\Gamma(\varpi_s^{\tau}-3)}{2} \varrho^{-1} \cos\pi k \\
    C_{\mathbf q, \tau} = \frac{\Gamma\varpi_s^{\tau}}{2} \varrho^{2}  \\
    D_{\mathbf q} = -2\tilde{J} \varrho^{-1} \cos\pi k \\
    E_{\mathbf q} = \frac{2\tilde{J} + \Gamma}{2} \varrho^{2}
  \end{array}
\right..
\end{align}
where $\varpi_s = \frac{1+2\sqrt2\imath}{3}$ and $\varrho = e^{\ii\pi h/3}$.

The quadratic Hamiltonian Eq.~\eqref{HtotAACCST} can be diagonalized via a bosonic Bogoliubov transformation\cite{SMJanssenVojta2019},
\begin{eqnarray}
	\left(\begin{array}{@{}cc@{}}
	\Omega(\mathbf q) & 0 \\
	0 & \Omega(\mathbf q)
	\end{array}\right)
	& = &
	T^\dagger(\mathbf q)
    \left(\begin{array}{cc}
	\hat{\Lambda}_{\bf q}              &    \hat{\Delta}_{\bf q}       \\
    \hat{\Delta}_{\bf q}^{\dagger}    &     \hat{\Lambda}_{-\bf q}^T
    \end{array}\right)
	T(\mathbf q)
\end{eqnarray}
where $\Omega(\mathbf q) = \diag{\omega_{\mathbf q,1}, \omega_{\mathbf q,2}, \cdots, \omega_{\mathbf{q}, n_s}}$.
The transformation matrix satisfies the orthogonality relations
$T \Sigma T^\dagger = T^\dagger \Sigma T = \Sigma$ where $\Sigma = \diag{\mathds 1, -\mathds 1}$.
The spectrum of the Hamiltonian can be obtained by the eigenvalue equation
\begin{eqnarray}
	\left(\begin{array}{cc}
	\hat{\Lambda}_{\bf q}              &    \hat{\Delta}_{\bf q}       \\
    -\hat{\Delta}_{\bf q}^{\dagger}    &     -\hat{\Lambda}_{-\bf q}^T
    \end{array}\right)
	\left(\begin{array}{@{}c@{}}
	\vec u^{(\upsilon)}_\mathbf{q} \\
	\vec v^{*(\upsilon)}_{-\mathbf q}
	\end{array}\right)	
	& = &
	\omega_{\mathbf{q},\upsilon}
	\left(\begin{array}{@{}c@{}}
	\vec u^{(\upsilon)}_\mathbf{q} \\
	\vec v^{*(\upsilon)}_{-\mathbf q}
	\end{array}\right).
\label{eq:eigenvectors-bogoliubov}
\end{eqnarray}
If the eigenvector $|n(\mathbf q) \rangle \equiv \left(\vec u^{(\upsilon)}_\mathbf{q}, \vec v^{*(\upsilon)}_{-\mathbf q}\right)^\mathrm{T}$ is normalized with respect to the inner product involving the matrix~$\Sigma$, i.e., $\langle n(\mathbf q) | \Sigma | n(\mathbf q)\rangle = 1$ with $\langle n(\mathbf q)| \equiv | n(\mathbf q)\rangle^\dagger$, then the columns of the matrix $T(\mathbf q)$ are given by the two vectors $|n(\mathbf q) \rangle$.

In addition to the spin-wave dispersion relations which are usually of prime interest,
there are also two other important quantities which can be easily obtained using the spin-wave calculation.
Namely, (i) the value of the total ordered moment $\langle M\rangle$ per site,
and (ii) the total energy per site $\varepsilon$.
For the classical moment $\langle M\rangle$, it is straightforwardly to get\cite{SMJanssenVojta2019}
\begin{eqnarray}
\frac{M}{S}
= \frac{1}{N_sn_s} \sum_{\{i\}\in N_s} \sum_{\{\upsilon\}\in n_s} \left(1 - \frac{1}{S} \langle a_{i,\upsilon}^\dagger a_{i,\upsilon} \rangle \right) + \mathcal O(1/S^2)
= 1 - \frac{1}{n_sS} \sum_{\{\upsilon\}\in n_s} \int \frac{d^2\mathbf q}{(2\pi)^2} \left|\vec v_{-\mathbf q}^{*(\upsilon)}\right|^2 + \mathcal O(1/S^2),
\end{eqnarray}
where $\vec v_{-\mathbf q}^{*(\upsilon)}$ denotes the lower half of the normalized $\upsilon$-th eigenvector occurring in Eq.~(\ref{eq:eigenvectors-bogoliubov}),
with positive energy $\omega_{\mathbf{q},\upsilon}$. The momentum integral is over all wavevectors $\mathbf q = (q_x, q_y)$ in the Brillouin zone.
Likewise, the spin wave energy $\varepsilon$ is given by\cite{SMJanssenVojta2019}
\begin{eqnarray}
\varepsilon = S(S+1) \varepsilon_{\textrm{cl}} + \frac{S}{2n_s} \sum_{\{\upsilon\}\in n_s} \int \frac{d^2\mathbf q}{(2\pi)^2} \omega_{\mathbf{q},\upsilon} + \mathcal O(1/S^2).
\end{eqnarray}

\subsection{two-sublattice AFM order.}
The AFM order has two sublattices in the honeycomb lattice.
Without loss of generality we can assume that $\tau = +1$ for the $\mathcal{A}$-sublattice and
$\tau = -1$ for the $\mathcal{B}$-sublattice.
Quite directly, we can obtain
\begin{eqnarray}\label{HtotAA}
\mathcal{H}
&=& -N_s S(S+1) \big[-\tilde{J} + 2\Gamma (\zeta^{y}\zeta^{z} + \zeta^{z}\zeta^{x} + \zeta^{x}\zeta^{y})\big] \nonumber \\
	&&+ \frac{S}{2} \sum_{\mathbf q}
	\left(\begin{array}{@{}c@{}}
	a_\mathbf{q} \\
    b_\mathbf{q} \\
	a_{-\mathbf{q}}^{\dagger} \\
    b_{-\mathbf{q}}^{\dagger}
	\end{array}\right)^\dagger
	\left(\begin{array}{@{}cccc@{}}
	\varepsilon_0           &   \lambda_0(\mathbf q)    &       0                   & \lambda_1(\mathbf q) \\
    \lambda_0^*(\mathbf q)  &   \varepsilon_0           &   \lambda_1(-\mathbf q)   & 0 \\
    0                       &   \lambda_1^*(-\mathbf q) &   \varepsilon_0           & \lambda_0^*(-\mathbf q)   \\
    \lambda_1^*(\mathbf q)  &   0                       &   \lambda_0(-\mathbf q)   & \varepsilon_0
	\end{array}\right)
	\left(\begin{array}{@{}c@{}}
    a_\mathbf{q} \\
    b_\mathbf{q} \\
	a_{-\mathbf{q}}^{\dagger} \\
    b_{-\mathbf{q}}^{\dagger}
	\end{array}\right)
\end{eqnarray}
where
\begin{align}
\left\{
  \begin{array}{l}
    \varepsilon_0  = -\tilde{J} + 2\Gamma (\zeta^{y}\zeta^{z} + \zeta^{z}\zeta^{x} + \zeta^{x}\zeta^{y})  \\
    \lambda_0(\mathbf q) = \Gamma \big[\mathcal{G}_{\tau=1}^{yz}(-1)e^{i{\mathbf q}\boldsymbol\delta_{x}}
        + \mathcal{G}_{\tau=1}^{zx}(-1)e^{i{\mathbf q}\boldsymbol\delta_{y}} + \mathcal{G}_{\tau=1}^{xy}(-1) e^{i{\mathbf q}\boldsymbol\delta_{z}}\big]  \\
    \lambda_1(\mathbf q) = -\tilde{J} (e^{i{\mathbf q}\boldsymbol\delta_{x}} + e^{i{\mathbf q}\boldsymbol\delta_{y}} - e^{i{\mathbf q}\boldsymbol\delta_{z}})
            + \Gamma \big[\mathcal{G}^{yz}(1)e^{i{\mathbf q}\boldsymbol\delta_{x}}
            + \mathcal{G}^{zx}(1)e^{i{\mathbf q}\boldsymbol\delta_{y}} + \mathcal{G}^{xy}(1)e^{i{\mathbf q}\boldsymbol\delta_{z}}\big]
  \end{array}
\right..
\end{align}

For the AFM ordering, the spins are found to be along the [111] direction.
So we choose the following crystalline axis,
$\mathbf{e} = \mathbf{a}\;[11\bar2]$, $\mathbf{n}\times\mathbf{e} = \mathbf{b}\;[\bar110]$, and $\mathbf{n} = \mathbf{c}^*\;[111]$.
In this case we have
\begin{align}\label{NaNb}
\big[\zeta_{\alpha}\zeta_{\beta}\big] = \frac13
  \left(
    \begin{array}{ccc}
      1 &  1 &  1 \\
      1 &  1 &  1 \\
      1 &  1 &  1 \\
    \end{array}
  \right),\quad
\big[\mathcal{G}^{\alpha\beta}(1)\big] =
  \frac13
  \left(
    \begin{array}{ccc}
      2 & -1 & -1 \\
     -1 &  2 & -1 \\
     -1 & -1 &  2 \\
    \end{array}
  \right), \quad
\big[\mathcal{G}_{\tau = 1}^{\alpha\beta}(-1)\big] =
  \frac23
  \left(
    \begin{array}{ccc}
     \omega^{-1}    &   1               & \omega        \\
     1              &   \omega          &  \omega^{-1}  \\
     \omega         &   \omega^{-1}     &  1            \\
    \end{array}
  \right)
\end{align}
where $\omega = e^{2\pi\ii/3}$.
In light of above equations we find that
\begin{align}
\left\{
  \begin{array}{l}
    \varepsilon_0  = 2\Gamma - \tilde{J} \\
    \lambda_0(\mathbf q) = 2\Gamma \gamma_{1,\mathbf{q}}  \\
    \lambda_1(\mathbf q) = -\tilde{J} \overline{\gamma}_{0,\mathbf{q}} - \Gamma \gamma_{0,\mathbf{q}}
  \end{array}
\right..
\end{align}
where
\begin{align}
\left\{
  \begin{array}{l}
    \overline{\gamma}_{0,\mathbf{q}} = e^{\ii{\mathbf{q}\boldsymbol\delta_{x}}} + e^{\ii{\mathbf{q}\boldsymbol\delta_{y}}} -e^{\ii{\mathbf{q}\boldsymbol\delta_{z}}}     \\
    \gamma_{0,\mathbf{q}} = \frac13\big(e^{\ii{\mathbf{q}\boldsymbol\delta_{x}}} + e^{\ii{\mathbf{q}\boldsymbol\delta_{y}}} +e^{\ii{\mathbf{q}\boldsymbol\delta_{z}}} \big)     \\
    \gamma_{1,\mathbf{q}} = \frac13\big(\omega^{-1} e^{\ii{\mathbf{q}\boldsymbol\delta_{x}}} + \omega e^{\ii{\mathbf{q}\boldsymbol\delta_{y}}} +e^{\ii{\mathbf{q}\boldsymbol\delta_{z}}} \big)
  \end{array}
\right..
\end{align}

\begin{figure}[!ht]
\centering
\includegraphics[width=0.95\columnwidth, clip]{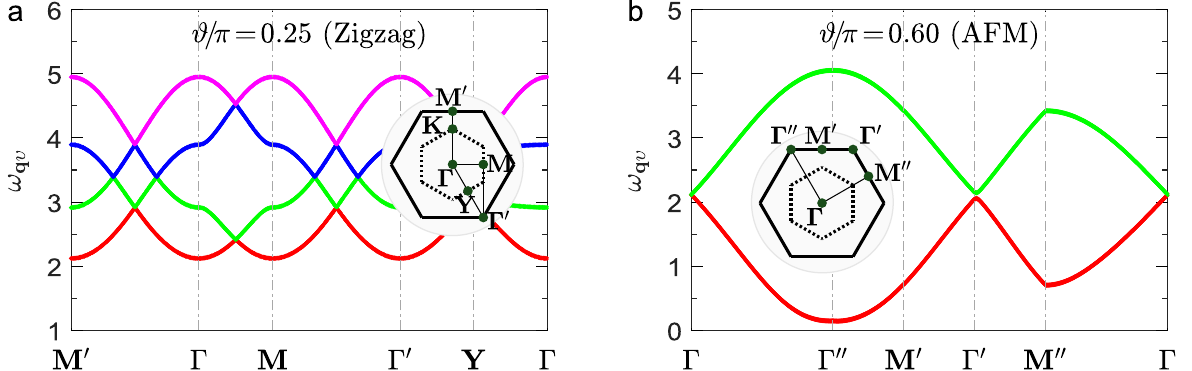}\\
\vspace{0.10cm}
\includegraphics[width=0.95\columnwidth, clip]{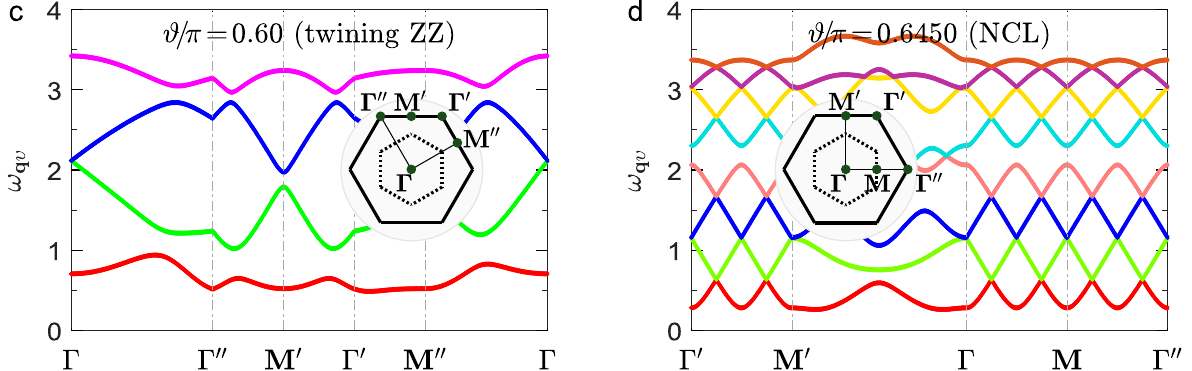}\\
\renewcommand{\figurename}{\textbf{Supplementary Figure}}
\caption{The spin-wave dispersions $\omega_{{\bf q}\upsilon}$ at $\vartheta/\pi = 0.25$ (zigzag), $\vartheta/\pi = 0.60$ (AFM),
    $\vartheta/\pi = 0.60$ (twining zigzag), and $\vartheta/\pi = 0.6450$ (NCL).
    The path in the reciprocal space is depicted in the inset for each subplot.}\label{FIGSM-DispABCD}
\end{figure}

\subsection{Spin-wave dispersion $\omega_{q\upsilon}$, energy $\varepsilon$, and sublattice magnetization $M$}

In this section, we begin by calculating the spin-wave dispersions for the selected points in
the zigzag phase ($0.00 \leq \vartheta/\pi < 0.50$),
the mixed phase of AFM ordering and twining zigzag ordering ($0.50 \leq \vartheta/\pi < 0.6368$),
and the noncolinear (NCL) phase ($0.6368 \leq \vartheta/\pi < 0.6543$).
The spin-wave dispersion of the stripy phase, which shows an order-by-disorder mechanism, is shown in the main text.
The spin-wave dispersions at $\vartheta/\pi = 0.25$ (zigzag), $\vartheta/\pi = 0.60$ (AFM), $\vartheta/\pi = 0.60$ (twining zigzag), and $\vartheta/\pi = 0.6450$ (NCL)
are shown in Fig.~\ref{FIGSM-DispABCD}, and the path in the reciprocal space is depicted in the inset for each subplot.
For the zigzag ordering (see Fig.~\ref{FIGSM-DispABCD}(a)), it is gapped with a magnon gap $\Delta$ of 2.12.
Generally, we have $\Delta = \sqrt{3(\Gamma^2+2\Gamma\tilde{J})}$ when $\vartheta \lesssim 0.40\pi$.
As can be seen from Fig.~\ref{FIGSM-DispABCD}(b) and (c), the magnon gap is relatively small
in the mixed phase of the AFM ordering and twining zigzag ordering.
In the NCL phase, its unit cell contains 8 sites (cf. Fig.~\ref{FIGSM-NonCop}(a)).
Therefore, the dispersion shown in Fig.~\ref{FIGSM-DispABCD}(d) contains eight branches and the minimal gap is around 0.25.

\begin{figure}[!ht]
\centering
\includegraphics[width=0.95\columnwidth, clip]{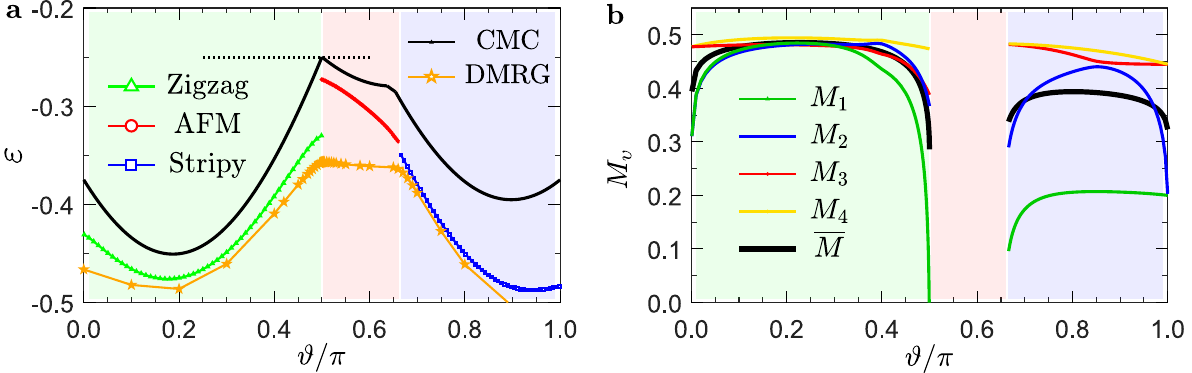}\\
\renewcommand{\figurename}{\textbf{Supplementary Figure}}
\caption{(a) The spin-wave energy for zigzag order (green), AFM order (red), and stripy order (blue).
    The black line is the classical energy ($S$ is set to be $1/2$) illustrated in the main text,
    and the yellow star line is the quantum energy on a 24-site hexagonal cluster.
    (b) The four branches of the sublattice magnetization $M_{\upsilon}$ ($\upsilon = 1-4$) for the zigzag order and stripy order.
    The averaged magnetization $\bar{M}$ (thick black line) is also shown for comparison.}\label{FIGSM-LSWTEgMag}
\end{figure}

We then turn to the spin-wave energy $\varepsilon$ and sublattice magnetization $M$ of the magnetically ordered states.
In Fig.~\ref{FIGSM-LSWTEgMag}(a),
the top black solid lines is the classical energy for all the phases.
The transitions between the neighboring phases are of first order because of the kinks in the energy curve.
The bottom yellow star line represents the quantum energy on a 24-site hexagonal cluster.
The spin-wave energy of the zigzag phase (green triangle), AFM phase (red circle), and stripy phase (blue square)
are calculated in their classically allowed region.
In the zigzag phase, the spin-wave energy correction $\Delta E$ (when compared to the classical energy)
is the largest at $\vartheta/\pi = 0.50$.
Here, the classical energy is -0.25 while the spin-wave energy is -0.33, showing an energy correlation of 0.08.
We note that the quantum ground-state energy at the $\Gamma$ limit is estimated to be -0.354(3) (see Fig.~\textcolor{red}{4} of the main text),
and there is a large quantum fluctuation due the classical ground-state degeneracy.
In contrast, in the intermediate region, the energy correction for the AFM phase is very small,
indicating that the AFM order is unlikely the true ground state at the quantum level.
We emphasize here that this phenomenon is directly related to the QSL phase by large-scale DMRG calculation.
In the stripy phase, the spin-wave energy is very close to the quantum result.

Figure~\ref{FIGSM-LSWTEgMag}(b) shows the sublattice magnetization $M_{\upsilon}$ ($\upsilon = 1-4$) of the zigzag and stripy phases.
Since both phases contain four sites in their unit cells, there are four different branches of $M_{\upsilon}$.
Deep into the zigzag phase, the four branches have very close values, but the difference becomes pronounced at the boundaries.
In the $\Gamma$ limit where $\vartheta/\pi \to 0.50$, the lowest branch $M_1$ is dramatically suppressed and it tends to vanish.
Since the quantum fluctuation has a strong impact on the lowest branch,
it thus suggests that the ground state of $\Gamma$ model is likely a QSL in the quantum situation.
However, we note that the averaged magnetization $\bar{M} = (M_1+M_2+M_3+M_4)/4$ is nonzero, with a value of $\bar{M} \approx 0.28$.
We will go back to this point later.
In the stripy phase, the lowest branch $M_1$ is relatively smaller than the remaining branches
and it is also insensitive to the interaction except near the phase boundary.
This phenomenon may relate to the fact the lowest branch is \textit{gapless} due to the emergent continuous symmetry.
In contrast, the second branch $M_2$ is more amenable to reveal the magnetization, and it share the same behavior as the quantum case.

\begin{figure}[!ht]
\centering
\includegraphics[width=0.75\columnwidth, clip]{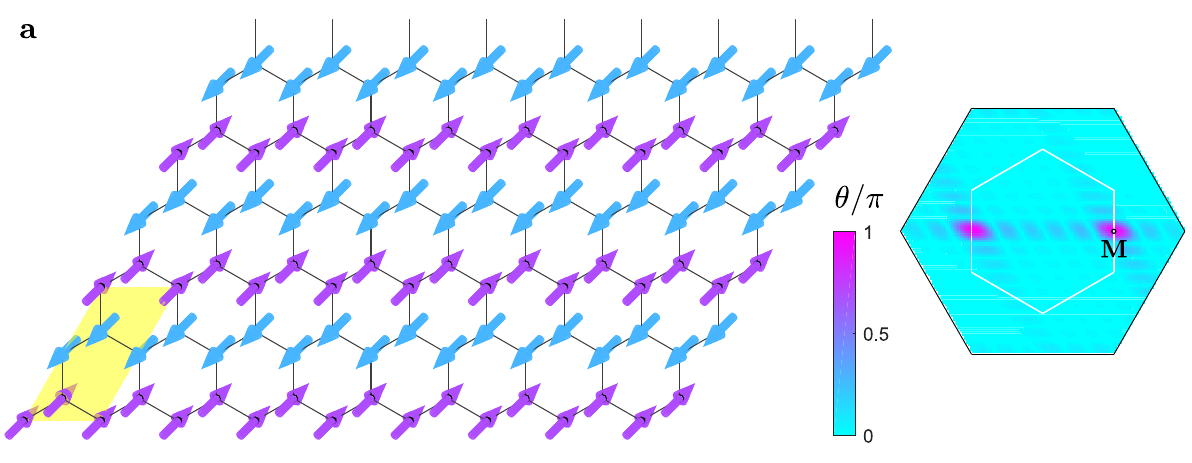}\\
\includegraphics[width=0.75\columnwidth, clip]{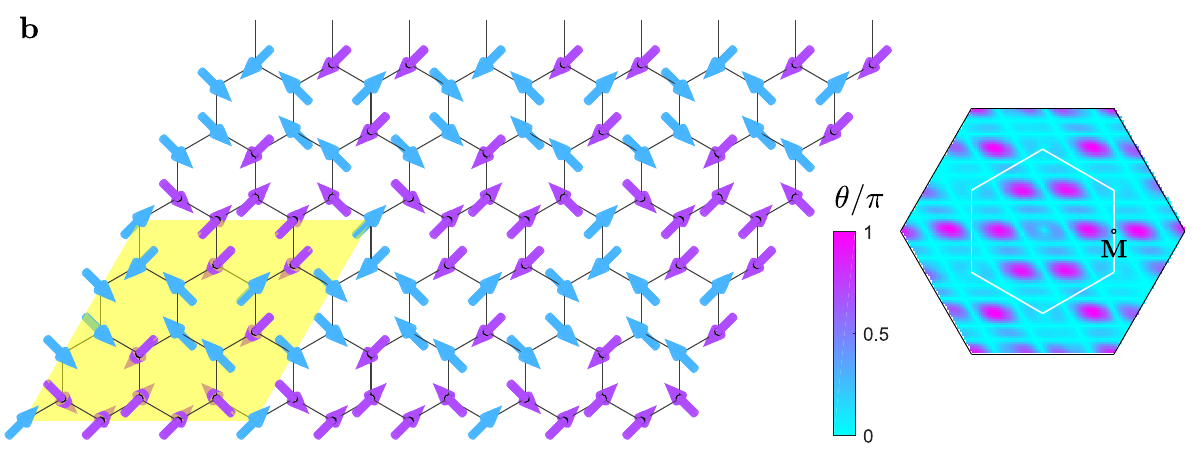}\\
\renewcommand{\figurename}{\textbf{Supplementary Figure}}
\caption{(a) Configuration of the zigzag ordering in the real space.
    The spins are parameterized by $\hat{S}=S\left(\sin{\theta}\cos{\phi},\sin{\theta}\sin{\phi},\cos{\theta}\right)$.
    Here, $\theta$ is represented by the color (see colormap) while $\phi$ is represented by the orientation of the arrow in the plane.
    The shape of the unit cell (with 4 sites in total) is marked by the yellow shadow.
    The right panel is the static structure factor in the momentum space with peaks at $\mathbf{M}$ points.
    (b) Configuration of one of the 18-site orderings in the real space.
    The shape of the unit cell (with 18 sites in total) is marked by the yellow shadow.
    The right panel is the static structure factor in the momentum space with peaks at $2\mathbf{M}/\mathbf{3}$ points.}
    \label{FIGSM-ZZvs18Site}
\end{figure}

Before ending this section, we wish to get further spin-wave signatures of the QSL in the pure $\Gamma$ model.
The classical $\Gamma$ model is known to have an infinite number of degeneracy,
with the two-sublattice AFM ordering, four-sublattice zigzag ordering, six-sublattice vortex-like 120$^{\circ}$ ordering,
and many large-unit-cell (LUC) ordering such as 18-site ordering.
In the quantum level, some of them are not favored (e.g., the AFM order, which has a higher spin-wave energy),
but a subset of them (whose number should be large) are competitive.
For example, the spin configurations of the zigzag ordering (peaks at $\mathbf{M}$ point) and 18-site ordering (peaks at $2\mathbf{M}/\mathbf{3}$ point)
are shown in Fig.~\ref{FIGSM-ZZvs18Site}.
For both of the two, all the spins have the same weight of three components, i,e., $|{S}_i^x|=|{S}_i^y|=|{S}_i^z|=S/\sqrt3$.
As a result, the zigzag ordering and the 18-site ordering have the same spin-wave energy and the same magnetization $\bar{M} \simeq 0.28$.
Since the zigzag ordering and the 18-site ordering have different peaks in the reciprocal space,
the linear superposition of the two leads to a reduced magnetization of ${\bar{M}}^\prime=\bar{M}/\sqrt2\approx0.20$.
In addition, by an alternative rearrangement of the spins, larger LUC orderings with 36-site and 48-site unit cells could be created,
and the superposition of them will further weaken the magnetization, giving rise to a nonmagnetic phase eventually.
In this sense, the spin-wave calculation is consistent with our large-scale DMRG calculation.

\vspace{0.50cm}
\begin{center}
\textsf{\textbf{Supplementary Note 3: The bond-modulated $\tilde{J}$-$\Gamma$ model: cylinder vs hexagonal cluster}}
\end{center}
\setcounter{subsection}{0}

\subsection{XC cylinder: precision, low-lying excitation, and entanglement entropy}

\begin{figure}[!ht]
\centering
\includegraphics[width=0.95\columnwidth, clip]{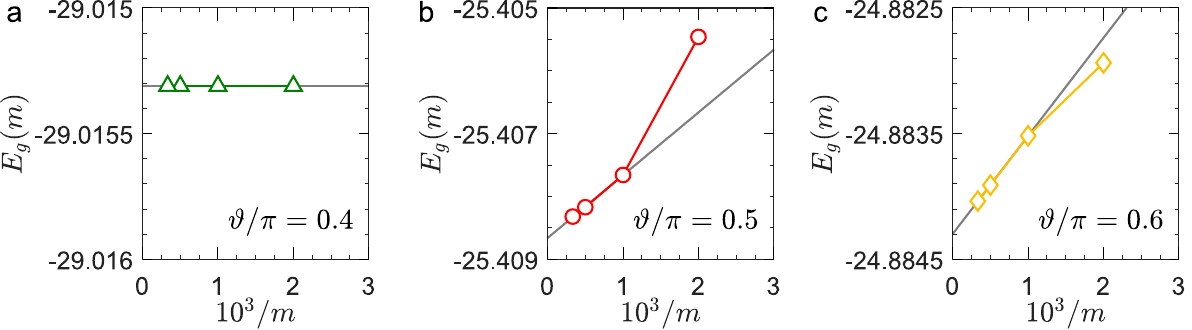}\\
\renewcommand{\figurename}{\textbf{Supplementary Figure}}
\caption{(a) Finite-size scaling of the total ground-state energy $E_g$ in the gapped zigzag phase ($\vartheta/\pi=\ 0.40$) with respect to $m$.
    (b) and (c) are for the intermediate phase at $\vartheta=0.50\pi$ (red circle) and $\vartheta=0.60\pi$ (yellow diamond).}\label{FIGSM-XCSweep}
\end{figure}

In the current computational capability, the number of block states $m$ in the two-dimensional (2D) DMRG calculation
is limited to a few thousands for the Hamiltonian without $U(1)$ symmetry.
Therefore, it is crucial to do a proper finite-size scaling of the measured quantities, such as energy and order parameter, with respect to $m$.
Taking the $12\times6$ XC cylinder as an example, we start by keeping $m$ = 500 states in the warming-up process.
Then we continue to sweep by increasing the block states to 1000, 2000, and 3000, respectively.
For each $m$ kept we perform 4 sweeps and up to 12 sweeps are applied.
The numbers of sweeps are extended to 16$\sim$20 times in the gapless region and the block state $m$ is increased to 4000 occasionally.
Figure~\ref{FIGSM-XCSweep} shows the extrapolation of the energy for the gapped zigzag phase ($\vartheta/\pi$= 0.40)
and the gapless QSL phase with $\vartheta/\pi$= 0.50 ($\Gamma$ limit) and $\vartheta/\pi$ = 0.60.
The energy in Fig.~\ref{FIGSM-XCSweep}(a) is almost $m$ independent, showing that the energy converges very quickly in the gapped phase.
In the QSL phase, the energy at $m$ = 1000, 2000, and 3000 shows a linear scaling of $1/m$.
Typically, the energy difference between the $E_g\left(m=3000\right)$ and the extrapolated one $E_g\left(m\ \rightarrow\ \infty\right)$ is not very big.
For example, when $\vartheta/\pi$= 0.50, the two are -25.40831921 and -25.40866029, respectively.
The energy per-site is -0.35289332 and -0.35289806, with an even smaller difference.

On the other hand, it is essential to check the ground-state degeneracy on XC cylinders,
the number of which could be large due to the possible gapless edge excitation.
To get the first few dozens of target states, we diagonalize a sparse Hermitian matrix by Davidson algorithm.
Suppose that we already have the first $k$ eigenvectors, we then extend the subspace and perform the iteration until some criterions are met,
and the $(k+1)$-th eigenvalue could be obtained with the default precision.
We emphasize that in the DMRG calculation, we need to use all the targeted states to construct the reduced density matrix.
In our implementation, all the target states are used with equal weight.
In principle, this method is general and can be used to target a large number of low-energy states.
But due to the truncation error, there should be a balance between the system sizes and the number of target states to guarantee the numerical accuracy.
For a cylinder with $\sim$50 sites, the first 20-30 states could be targeted precisely with an error bar of $\mathcal{O}\left({10}^{-6}\right)$ or less.

\begin{figure}[!ht]
\centering
\includegraphics[width=0.45\columnwidth, clip]{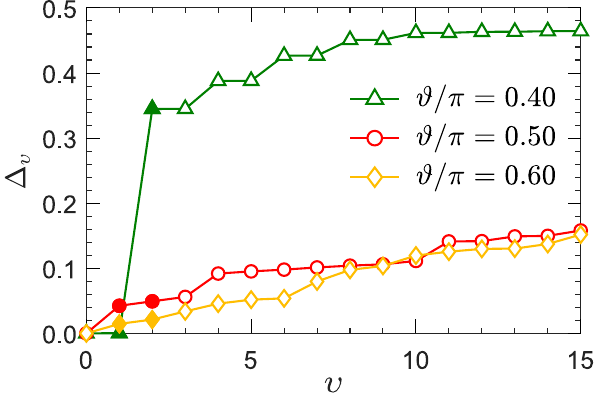}\\
\renewcommand{\figurename}{\textbf{Supplementary Figure}}
\caption{The first fifteen energy gap $\Delta_{\upsilon} = E_{\upsilon} - E_0$ on a XC cylinder of $8\times4$.
    The underlying phases are gapped zigzag phase ($\vartheta=0.40\pi$, green triangle),
    and intermediate phase at $\vartheta=0.50\pi$ (red circle) and $\vartheta=0.60\pi$ (yellow diamond).}\label{FIGSM-XC15Gap}
\end{figure}

In our calculation, we calculate the first sixteen energy levels on a small system size of $8\times4$ cylinder,
which enables us to get the low-lying energy accurately.
The excitation gap $\Delta_{\upsilon} = E_{\upsilon} - E_0$ with $\upsilon = 0-15$ is shown in Fig.~\ref{FIGSM-XC15Gap}.
For the zigzag order ($\vartheta=0.4\pi$), the ground state is indeed doubly degenerate with a large energy gap of $\sim0.35$
(Note: we note that this value is somewhat small due to small system size and cylinder boundary condition.
As shown in Fig. \textcolor{red}{6}a in the main text, excitation gap for larger cylinder is $\sim0.65$,
which is consistent with the PBC case shown in Tab. \textcolor{red}{1} in the main text).
For the intermediate phase at $\vartheta=0.5\pi$ and $\vartheta=0.6\pi$, the excitation gap is successive increasing without large abrupt change.
Thus, there is unlikely a large number of ground-state degeneracy.

\begin{figure}[!ht]
\centering
\includegraphics[width=0.45\columnwidth, clip]{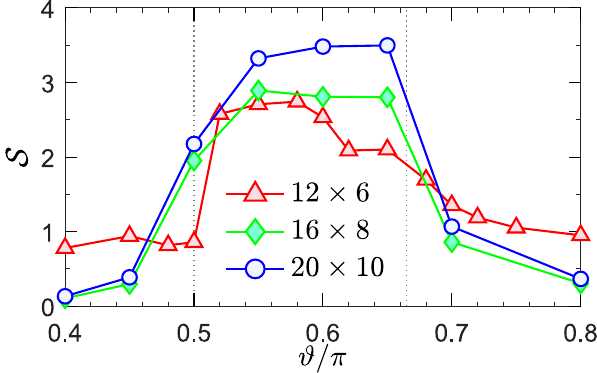}\\
\renewcommand{\figurename}{\textbf{Supplementary Figure}}
\caption{The von Neumann entropy $\mathcal{S}$ on XC clusters of $12\times6$ (red triangle), $16\times8$ (green diamond),
    and $20\times10$ (blue circle).}\label{FIGSM-XCVNE}
\end{figure}

In Fig.~\textcolor{red}{4} of the main text, we show the ground-state energy
on XC clusters of $12\times6$, $16\times8$, and $20\times10$.
Here, Fig.~\ref{FIGSM-XCVNE} show the von Neumann entropy $\mathcal{S}$ on the same cylinders.
It is observed that the entropy in the gapped zigzag and stripy phases are lower than that in the intermediate phase.
With the increasing of the total sites, the entropy in the intermediate phase increases,
consistent with the conclusion that the intermediate phase is gapless.
In addition, the entropy has drops at $\vartheta_{l}/\pi \simeq 0.50$ and $\vartheta_{r}/\pi \approx 0.66$,
indicating that both of the transitions are of first order.

\subsection{Hexagonal cylinder: Zigzag-QSL transition}

\begin{figure}[!ht]
\centering
\includegraphics[width=0.85\columnwidth, clip]{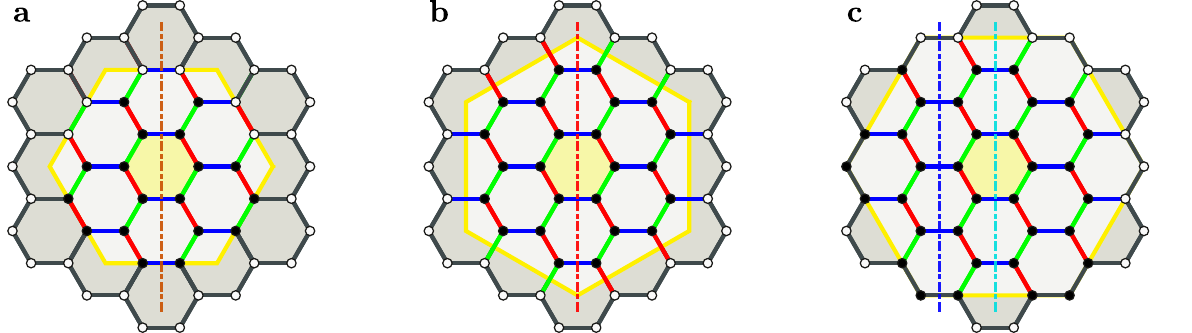}\\
\renewcommand{\figurename}{\textbf{Supplementary Figure}}
\caption{The lighted regions with yellow frames mark hexagonal clusters of (a) $N$ = 18, (b) $N$ = 24, and (c) $N$ = 32.
    The black solid dots represent real sites whose number should be $N$.
    The types of bonds are \textbf{X}-bond (red), \textbf{Y}-bond (green), and \textbf{Z}-bond (blue).
    In (a) and (b), the dotted vertical lines represent the partition of the systems into two equal halves with $N/2$ sites in each part.
    In (c), the blue and cyan dotted lines represent $[l, r]$-partitions of [11, 21] and [19, 13], respectively, with four \textbf{Z}-bonds been cut.}\label{FIGSM-Hex182432}
\end{figure}

Due to the bond-modulated Heisenberg ($\tilde{J}$) interaction which favors the zigzag ordering,
the Zigzag-QSL transition near $\vartheta_l/\pi\ \simeq 0.50$ is more intricate.
To determine the transition type and the transition point accurately,
we have done extra calculation on hexagonal clusters of $N$ = 18, 24, and 32 under full periodic conditions (for geometries, see Fig.~\ref{FIGSM-Hex182432}).
We note that although these clusters have less sites than the cylinder cases,
they do not have the boundary effect and thus show clearer tendencies of physical quantities as $N$ is increased.
To begin with, we calculate the energy and von Neumann entropy on the 18-site cluster.
It is found that there is a kink in the energy $e_g$ and a sharp jump in the entropy $\mathcal{S}$ (not shown), supporting the first-order transition.
Nevertheless, since the 18-site cluster does not match with the zigzag ordering whose unit cell is 4, this result is more or less unjust.

\begin{figure}[!ht]
\centering
\includegraphics[width=0.85\columnwidth, clip]{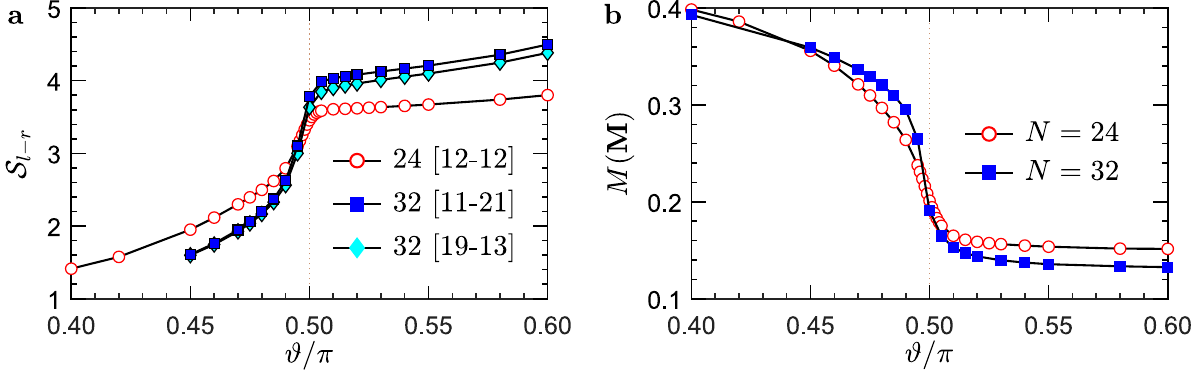}\\
\renewcommand{\figurename}{\textbf{Supplementary Figure}}
\caption{(a) The von Neumann entropy on 24-site (open symbol) and 32-site (fill symbols) hexagonal clusters.
    The partitions of the systems are shown in Fig.~\ref{FIGSM-Hex182432}(b) and -\ref{FIGSM-Hex182432}(c).
    (b) The order parameter of the zigzag ordering on the 24-site (red circle) and 32-site (blue square) clusters.}\label{FIGSM-HexVNEJump}
\end{figure}

In what follows, we resort to the 24-site and 32-site clusters, which are accommodative with the zigzag phase.
First of all, since the entropy is amenable to signify the transition type,
we thus calculate the von Neumann entropy $\mathcal{S}$ by cutting the same number of bonds.
As can be seen from Fig.~\ref{FIGSM-Hex182432}(b) and \ref{FIGSM-Hex182432}(c),
the red cut in the 24-site case and the blue and cyan cuts in the 32-site case meet this requirement where four \textbf{Z}-Type bonds are cut,
and the entropy are shown in Fig.~\ref{FIGSM-HexVNEJump}(a).
In the transition region where $0.48 < \vartheta/\pi < 0.51$, the entropy for the 24-site cluster increases from 2.4984 to 3.6065, an increment of 1.1081.
By contrast, the increment is $\sim$1.80 for the 32-site case.
This indicates that the difference of the entropy in the zigzag phase and the QSL phase should be enlarged as $N$ is increased,
and there should be a jump at the transition point as $N \rightarrow \infty$.
In addition, we also calculate the order parameter $M(\textbf{M})$ of the zigzag ordering for the two system sizes.
As can be seen in Fig.~\ref{FIGSM-HexVNEJump}(b), the order parameter has a tendency to jump with the increasing of $N$.
In summary, the results of entropy and order parameter both favor the first-order Zigzag-QSL transition.

\begin{figure}[!ht]
\centering
\includegraphics[width=0.90\columnwidth, clip]{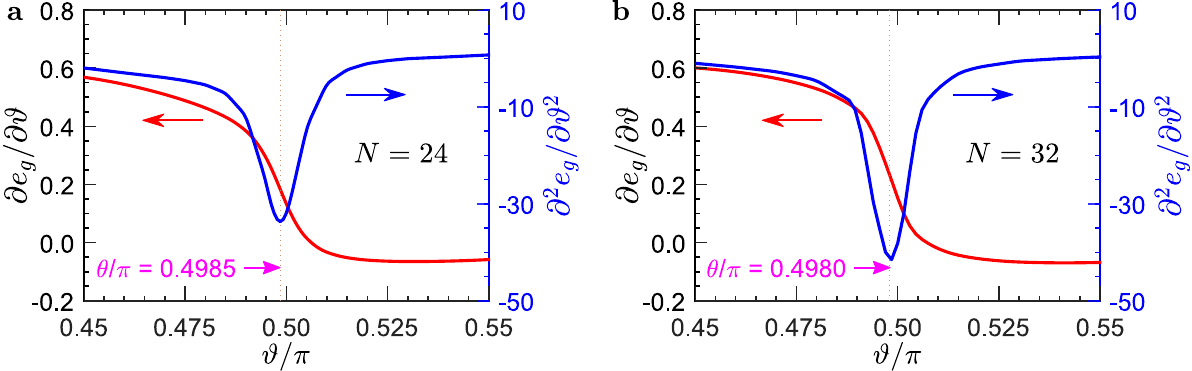}\\
\renewcommand{\figurename}{\textbf{Supplementary Figure}}
\caption{(a) The first-order (in red, left) and second-order (in blue, right) energy derivatives of a 24-site hexagonal cluster.
    The peak position locates at $\vartheta/\pi$ = 0.4985.
    (b) The same as (a) but for a 32-site cluster. The peak position locates at $\vartheta/\pi$ = 0.4980.}\label{FIGSM-HexEgDerPeak}
\end{figure}

In the main text, we estimate that the transition point $\vartheta_{l}/\pi$ is around but slightly smaller than 0.50.
It is the staggered-like Heisenberg ($\tilde{J}$) interaction that enhances the zigzag phase and pushes the transition point to the $\Gamma$ limit.
To make a reasonable and precise transition point, we turn to the energy derivatives on hexagonal clusters of $N$ = 24 and $N$ = 32.
For a \textit{finite-size} system, there is a peak in the second-order energy derivative.
As the system size is increased, the peak will behave as a Gaussian-like wave-packet (~Dirac $\delta$ function) for the first-order transition
or will diverge for continuous phase transition.
The ground-state energy is calculated with an increment of 0.001 and 0.005 $\left(\vartheta/\pi\right)$ for the $N$ = 24 and 32 cases, respectively.
The energy is extrapolated by a spline-extrapolation method so as to make the energy derivatives be smooth.
The first-order (red) and second-order (blue) energy derivatives are shown in Fig.~\ref{FIGSM-HexEgDerPeak}.
For the $N$ = 24 case (left panel), the peak locates at 0.4985, while it is 0.4980 for the 32-site case (right panel).
The transition point is slowly shifted to the left and we expect it will not change too much as $N$ is further increased.
Thus, we estimate the transition point as $\vartheta/\pi=0.498\left(1\right)$,
and the pure $\Gamma$ limit falls in the intermediate phase.

\vspace{0.50cm}
\begin{center}
\textsf{\textbf{Supplementary Note 4: The honeycomb $\Gamma$ model: energy gap scaling and entanglement entropy scaling}}
\end{center}
\setcounter{subsection}{0}

\subsection{energy gap scaling}

In a recent study of the one-dimensional (1D) $\Gamma$ chain by us\cite{SMLuoKGarXiv2020},
the ground-state energy $e_g$ has been shown to have an intimate relation to the boundary condition.
Notably, $e_g$ exhibits a six-site periodicity with the length of $L_x$.
We recall that the gap for vison excitations in the Kitaev honeycomb model shows
a three-period structure, as pointed out by Kitaev\cite{SMKitaev2006}.
We thus speculate that the energy spectrum of the two-dimensional (2D) $\Gamma$ model
should also own a similar but more complicated periodicity.

\begin{figure}[!ht]
\centering
\includegraphics[width=0.55\columnwidth, clip]{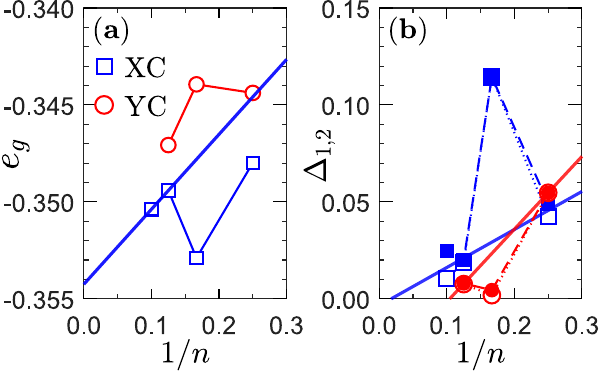}\\
\renewcommand{\figurename}{\textbf{Supplementary Figure}}
\caption{(a) Ground-state energy $e_{g}$ of $\Gamma$ model on XC (blue) and YC (red) clusters.
    The system sizes $L_x\times L_y$ are $8\times4$ ($n = 4$), $12\times6$ ($n = 6$), $16\times8$ ($n = 8$), and $20\times10$ ($n = 10$).
    (b) Energy gaps $\Delta_{1}$~(open symbols) and $\Delta_{2}$~(filled symbols) for the corresponding clusters.
    The thick solid lines indicate the overall trends of the curves.}\label{FIGSM-GamGap}
\end{figure}

The unusual size-dependent behavior has an awful impact on the energy and energy gaps
shown in Fig.~\ref{FIGSM-GamGap},
making an accurate extrapolation to the thermodynamic limit intractable.
We have calculated the low-lying energy on both XC clusters (blue square) of $8\times4$, $12\times6$, $16\times8$, and $20\times10$,
and also YC clusters (red circle) of $8\times4$, $12\times6$, and $16\times8$, see Fig.~\ref{FIGSM-GamGap}.
The XC cluster shown in Fig.~{\textcolor{red}{1}} of the main text has zigzag open edges,
while the YC cluster is a $90^{\circ}$ rotation of the XC cluster and has armchair open edges.
In both cases, the energy has an oscillation, and we estimate the ground-state energy is -0.354(3)
(more details are presented in Fig.~{\textcolor{red}{4}} of the main text).
As shown in Fig.~\ref{FIGSM-GamGap}(b), there are similar oscillations in the energy gaps $\Delta_1$ (open) and $\Delta_2$ (filled).
Because of the overall downward trend in the gaps with the increasing of the system size,
we estimate that the lowest gap should be $\Delta = 0.00(1)$ in the thermodynamic limit.
This result is in line with the dense energy spectrum on a 24-site hexagonal cluster shown in Fig.~{\textcolor{red}{6}a} of the main text),
favoring the gapless ground state in the $\Gamma$ model.

Next, we go beyond the XC/YC cylinders with a fixed $L_x/L_y = 2$,
and study the energy gap of the $\Gamma$ model ranging from 1D chain to two-leg honeycomb ladder,
and also towards a series of $2\times L_x\times L_y$ tori of $L_y$ = 3 or 4.
The geometry of the latter is shown in the inset of Fig.~\textcolor{red}{7} in the main text.

\begin{figure}[!ht]
\centering
\includegraphics[width=0.50\columnwidth, clip]{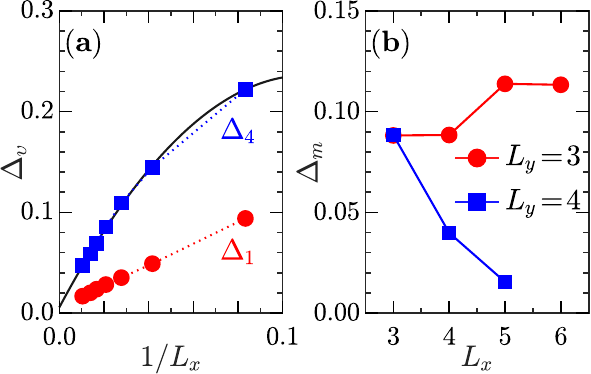}\\
\renewcommand{\figurename}{\textbf{Supplementary Figure}}
\caption{(a) Energy gap $\Delta_{\upsilon} = E_{\upsilon} - E_0$ for a two-leg honeycomb ladder under PBC.
    The black line is the extrapolation of $\Delta_4$ to infinite-size system.
    (b) The minimal energy gap $\Delta_m$ for the three-leg (red circle) and four-leg (blue square) tori.}\label{FIGSM-LTGap}
\end{figure}

\begin{itemize}
  \item To begin with, for the 1D isotropic $\Gamma$ chain,
    its ground state is found to be a gapless Luttinger liquid\cite{SMLuoKGarXiv2020,SMYangKG2020} with emergent $SU(2)$ symmetry.
  \item Furthermore,  we consider a two-leg honeycomb ladder which is a rung-alternating coupling of two isotropic $\Gamma$ chain.
    It is a stripe of a honeycomb lattice along its zigzag edges and only contains $L_x/2$ $\textrm{Z}$ bonds.
    We find that there is a unique ground state with energy $E_0$ under PBC, followed by a triplet excited state with energy $E_1$. There seems to be a continuous spectrum afterwards and the lowest branch has a energy of $E_4$.
    Figure~\ref{FIGSM-LTGap}(a) shows the energy gap of $\Delta_1 = E_1-E_0$ and $\Delta_4 = E_4-E_0$, which go down as $L_x$ increased.
    After an extrapolation of $\Delta_4$ we find that $\Delta_4 < 0.004$, which seems to close for long enough ladder.
  \item Moreover, we study the energy gaps for three- and four-leg tori.
    Here, PBCs are imposed on both directions so as to remove the possible edge excitations.
    For $L_y = 3$, we perform the calculation on four different tori with $L_x$ = 3, 4, 5, and 6, and find that the gap is around $\sim0.11$.
    However, for $L_y = 4$, the gap goes down quickly from 0.09 when $L_x = 3$ to 0.015 when $L_x = 5$, see Fig.~\ref{FIGSM-LTGap}(b).
    We thus infer that the gap for $L_y = 4$ most probably vanishes as $L_x \to \infty$.
\end{itemize}

The results on different clusters are summarized in the table below.
We find that the energy gap has a strong cluster dependence,
and could vanishes at several cases.
We note that this is not the typical character of a gapped system whose gap is usual very stable.
In this regard, it is another evidence for the gaplessness of $\Gamma$ model.

\begin{table}[th!]
\renewcommand{\tablename}{\textbf{Supplementary Table}}
\caption{\label{Tab-ChiEv}
Energy gap of the pure $\Gamma$ model under a 1D chain, two-leg honeycomb ladder,
and also $2\times L_x\times L_y$ tori of $L_y$ = 3 or 4.}
\begin{tabular}{ c | c | c}
\colrule\colrule
Cases                               & Energy gap        & Gapped/Gapless    \\
\colrule
1D isotropic $\Gamma$ chain         & 0                 & gapless   \\
two-leg honeycomb $\Gamma$ ladder   & $<0.004$          & gapless   \\
$2\times L_x\times3$ torus          & $\sim0.11$        & gapped    \\
$2\times L_x\times4$ torus          & $<0.001$          & gapless   \\
\colrule\colrule
\end{tabular}
\end{table}

\subsection{entanglement entropy scaling}

\begin{figure}[!ht]
\centering
\includegraphics[width=0.50\columnwidth, clip]{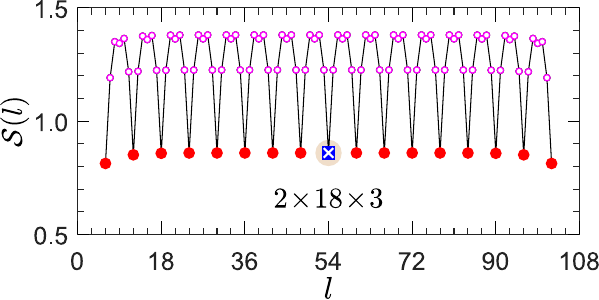}\\
\renewcommand{\figurename}{\textbf{Supplementary Figure}}
\caption{Entanglement entropy $\mathcal{S}(l)$ of a consecutive segment of length $l$ on a $2\times18\times3$ cylinder.
    The solid symbols of the lowest branch represent the neat edge-cutting with $l$ being a multiply of 6 (i.e., the number of the sites along each column).
    The bipartite entanglement entropy with $l = N/2$ is marked as a blue square.}\label{FIGSM-VNE1806}
\end{figure}

As shown in the last subsection, the three-leg cylinder is found to be gapped in the pure $\Gamma$ model.
Such a gapped ground state could also be checked by the entanglement entropy.
Figure~\ref{FIGSM-VNE1806} shows the representative behavior of $\mathcal{S}(l)$ on a $2\times18\times3$ cylinder,
which contains six sites along each column. When $l$ is a multiply of 6, it corresponds to a neat edge-cutting where the two halves have smooth margins.
The entanglement entropy is minimized and forms a lower branch as marked by solid red symbols.
It is clearly found that the lower branch is very flat in the middle region,
in accordance with a gapped system with a central charge of 0.

\begin{figure}[!ht]
\centering
\includegraphics[width=0.95\columnwidth, clip]{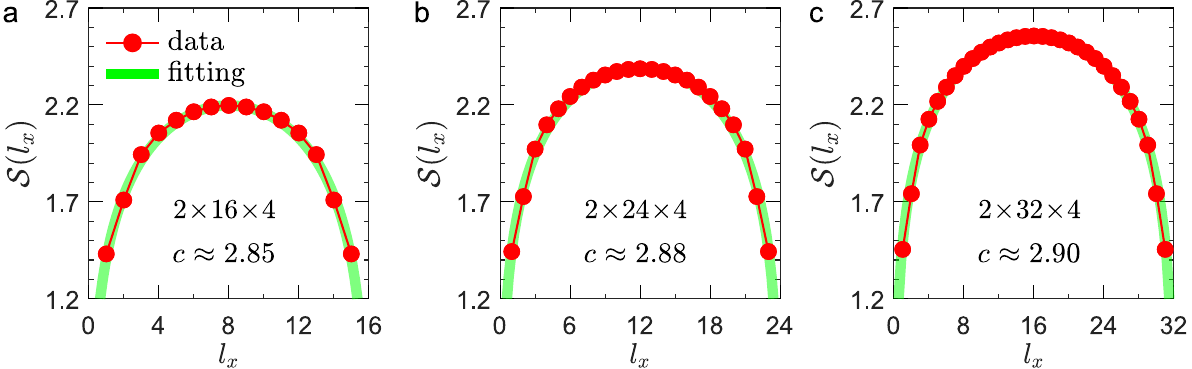}\\
\renewcommand{\figurename}{\textbf{Supplementary Figure}}
\caption{(a) Extracting the central charge $c$ from the entanglement entropy $\mathcal{S}$ on a $2\times16\times4$ cylinder.
    The fitting formula is Eq.~\eqref{EQSM:VNECC} and the central charge is estimated as 2.85.
    (b) and (c) show the entanglement entropy scaling on cylinders of $2\times24\times4$ and $2\times32\times4$, respectively,
    with the fitting central charge $c \approx 2.88$ and $c \approx 2.90$.}\label{FIGSM-VNELy04CC}
\end{figure}

By contrast, the four-leg cylinder is found to be gapless.
For the quasi-one-dimensional conformal invariant critical system under open boundary condition,
it is established that the entanglement entropy obeys the following formula \cite{SMEisert2010}
\begin{equation}\label{EQSM:VNECC}
\mathcal{S}(l_x) = \frac{c}{6}\ln\left[\frac{2L_x}{\pi}\sin\Big(\frac{\pi l_x}{L_x}\Big)\right] + c'
\end{equation}
where $c$ is the central charge and $c'$ is a model-dependent fitting constant.
$l_x$ is the number of the columns of the subsystem and $L_x$ is the length of the cylinder.
In Fig.~\textcolor{red}{7} of the main text, we take $l_x = L_x/2$ and fit the central charge as
$\mathcal{S} = \frac{c}{6}\ln\left(\frac{2L_x}{\pi}\right) + c'$.
In that case we find that $(c, c') \approx (2.92, 1.09)$, showing that the central charge is approximately 3.
Here, instead, we turn to fit the central charge on each individual cylinder according to the formula Eq.~\eqref{EQSM:VNECC}.
Figure~\ref{FIGSM-VNELy04CC} shows the fitting of central charge on cylinders of $2\times16\times4$, $2\times24\times4$, and $2\times32\times4$.
We find that the best fitting values are 2.85, 2.88, and 2.90, respectively.
Therefore, the central charge converges to 3 for long enough four-leg cylinder.

\begin{figure}[!ht]
\centering
\includegraphics[width=0.50\columnwidth, clip]{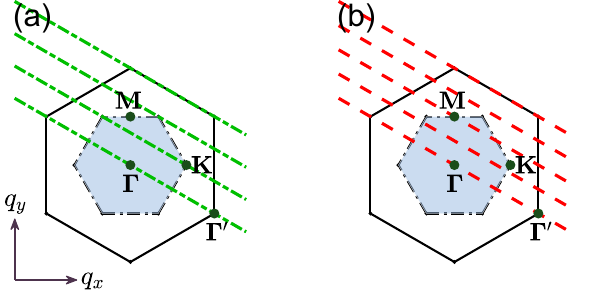}\\
\renewcommand{\figurename}{\textbf{Supplementary Figure}}
\caption{Quantized momenta along the circumferences of the cylinders of $L_y = 3$ (left) and $L_y = 4$ (right).}\label{FIGSM-BZLine}
\end{figure}

In the end, we want to address that the significant difference between $L_y$ = 3 and 4 suggests an unusual way from multi-leg ladder towards 2D limit.
Besides, it may imply that the ground state is likely to own spinon Fermi surface (SFS) \cite{SMJiangWHetalarXiv2018,SMPatelTrivedi2019}.
In this scenario, the pockets of SFS could be detected by different cuts along the Brillouin zones.
Since the pockets are usually distributed at several high symmetry points,
the central charge should vary for different width of cylinders, depending on how many pockets are crossed.
Such a $L_y$-dependent behavior of the central charge is a highlight signature of SFS QSL.
Figure~\ref{FIGSM-BZLine} shows the quantized momenta along the circumference
of the three-leg (left) and four-leg (right) cylinders.
In view of the different central charges of the two cases,
we speculate the gapless excitations meet red lines several times in Fig.\ref{FIGSM-BZLine}(b)
but not the green lines in Fig.~\ref{FIGSM-BZLine}(a).
However, since there is no translational symmetry along the $L_x$-direction (which is OBC),
we cannot determine the precise momenta of the gapless excitations.

Nevertheless, the central charge $3$ on four-leg cylinders may also match with the possibility of a Dirac QSL
where three Dirac Fermions are located around $\textbf{M}$ points.
We note that central charge on wider cylinders of $L_y = 5$ and 6 should be useful to further clarify this issue.
Actually, in the finite-size DMRG calculation, the width of a long cylinder ($L_x\lesssim50$) is usually limited to three or four unit-cell,
as the precision is not satisfactory for wider cylinders.
So it is unclear currently how does the central charge vary with further increasing of width $L_y$.
We speculate that this problem might be studied by the variational Monte Carlo calculation
on a mean-field Hamiltonian constructed out of Abrikosov fermions or Majorana fermions \cite{SMIqbalBSP2013}
or the infinite DMRG calculation on the correlation length spectrum \cite{SMHuZEH2019}.
We think that this is an exciting research direction in the future.

\vspace{0.50cm}
\begin{center}
\textsf{\textbf{Supplementary Note 5: Plaquette order parameter}}
\end{center}
\setcounter{subsection}{0}

In the Kitaev honeycomb model, the hexagonal plaquette operator $\hat{W}_p = 2^6 S_1^{x}S_2^{y}S_3^{z}S_4^{x}S_5^{y}S_6^{z}$
commutates with the model and $\hat{W}_p = \pm 1$\cite{SMKitaev2006}.
For the $\Gamma$ model as well as the general bond-modulated $\tilde{J}$-$\Gamma$ model, $[\mathcal{H}, \hat{W}_p] \neq 0$,
so $\hat{W}_p$ is no longer a conserved quantity.
However, the flux-like density $\langle \overline{W}_p\rangle = \sum_p \langle \hat{W}_p\rangle/N_p$
where $N_p = N/2$ is the number of hexagonal plaquette
can be tremendously useful and informative to distinguish different phases.

\begin{figure}[!ht]
\centering
\includegraphics[width=0.50\columnwidth, clip]{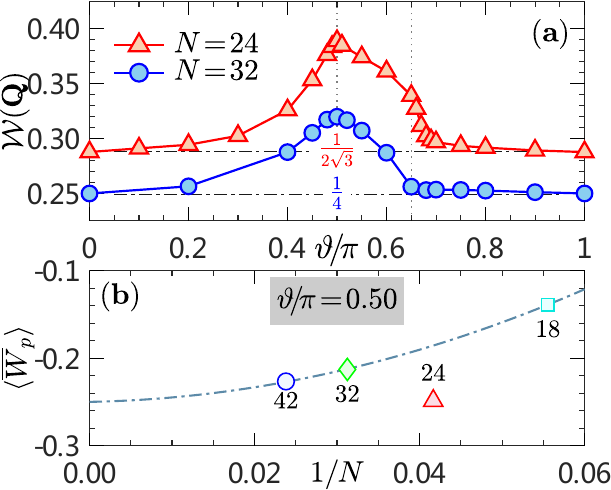}\\
\renewcommand{\figurename}{\textbf{Supplementary Figure}}
\caption{(a) SVSF $\mathcal{W}(\textbf{K})$ on hexagonal clusters of $N = 24$ and 32.
The horizonal lines of $1/(2\sqrt3)$ and $1/4$ are approximately the lower limits at the corresponding sizes.
(b) Extrapolated flux-like density $\langle \overline{W}_p\rangle$ for $\Gamma$ magnet.}\label{FIGSM-SVSF}
\end{figure}

We calculate the plaquette-plaquette correlation $\langle \hat{W}_p \hat{W}_q\rangle$,
and define the static plaquette structure factor\cite{SMSahaFZetal2019},
\begin{equation}\label{EQSM-VisonSF}
\mathcal{W}_{N_p}({\bf{Q}})=\frac{1}{N_p}\sum_{pq}
\langle{\hat{W}_p \hat{W}_q}\rangle e^{i{\bf{Q}}\cdot{({\bm{R}}_p-{\bm{R}}_q)}},
\end{equation}
where $\bm{R}_p$ is the central position of each plaquette.
We find that there is a dominating peak in the $\boldsymbol{\Gamma}$ point
and also a subleading peak at $\bf{K}$ point of the Brillouin zone.
This fact implies that there is no translational symmetry breaking in the honeycomb lattice
but with perceptible plaquette correlation.
We thus define the plaquette order parameter as
$\mathcal{P}_{N_p} = \sqrt{\mathcal{W}_{N_p}({\bf{Q}})/N_p}$.
The results on the hexagonal clusters of $N = 24$ and 32 are shown in Fig.~\ref{FIGSM-SVSF}(a).
Due to the dominating contribution from the trivial identity $\big\langle (\hat{W}_p)^2\big\rangle$ = 1
\footnote{Since $\langle (S^x)^2\rangle = \langle (S^y)^2\rangle = \langle (S^z)^2\rangle = 1/4$, we can get
$\big\langle (W_p)^2\big\rangle$ = $2^{12}\big(\langle (S_{\upsilon}^{\alpha})^2\rangle\big)^6 = 2^{12}\cdot\big(1/4^6\big) = 1$.},
$\mathcal{P}_{N_p}$ has a considerable finite-size value, which is approximately $1/\sqrt{N_p}$,
see the horizonal lines in Fig.~\ref{FIGSM-SVSF}.
Therefore, the bared plaquette order parameter is formally defined as
\begin{equation}\label{EQSM-HiddenOPPlqtt}
\mathcal{P}_{N_p} = \sqrt{\frac{\mathcal{W}_{N_p}({\bf{Q}})}{N_p}} - \frac{1}{\sqrt{N_p}}.
\end{equation}

The result of the bared plaquette order parameter is shown in Fig.~\textcolor{red}{7} of the main text.
It can be found that there is a decreasing plaquette correlation around $\Gamma$ model.
Besides, there is a negative flux-like density $\langle \overline{W}_p\rangle = -0.25(2)$ for $\Gamma$ model (see Fig.~\ref{FIGSM-SVSF}(b)).

\vspace{0.50cm}
\begin{center}
\textsf{\textbf{Supplementary Note 6: Role of 3rd-NN interaction}}
\end{center}
\setcounter{subsection}{0}

\begin{figure}[!ht]
\centering
\includegraphics[width=0.60\columnwidth, clip]{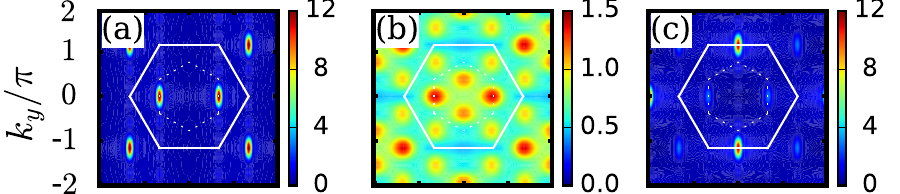}\\
\includegraphics[width=0.60\columnwidth, clip]{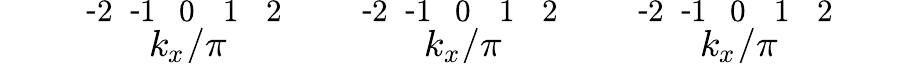}\\
\renewcommand{\figurename}{\textbf{Supplementary Figure}}
\caption{Typical contour plots of the overall SMSF for
(a)~zigzag phase~($\vartheta/\pi = 0.25$),
(b)~QSL phase~($\vartheta/\pi = 0.50$), and
(c)~stripy phase~($\vartheta/\pi = 0.75$)
are shown for XC clusters of $12\times6$.}\label{FIGSM-MSFPD}
\end{figure}

For the quantum bond-modulated $\tilde{J}$-$\Gamma$ model, there are three distinct phases,
including a zigzag order, a stripy order, and also a QSL.
The ground state of $\Gamma$ model belongs to the QSL phase but locates very close to the transition point between the zigzag order and the QSL.
The selected contour plots of the SMSF for the three phases are shown in Fig.~\ref{FIGSM-MSFPD}(a)-(c).
While the zigzag and stripy phases peak at $\textrm{\bf{M}}$ and/or $\textrm{\bf{M}}'$ points,
the magnetic order at $\vartheta/\pi = 0.5$ is tiny,
and a subleading peak locating at $\textrm{\bf{X}}$ point in the Brillouin zone appears.
This peak could be enhanced by negative third-NN $J_3$ interaction.

\begin{figure}[!ht]
\centering
\includegraphics[width=0.50\columnwidth, clip]{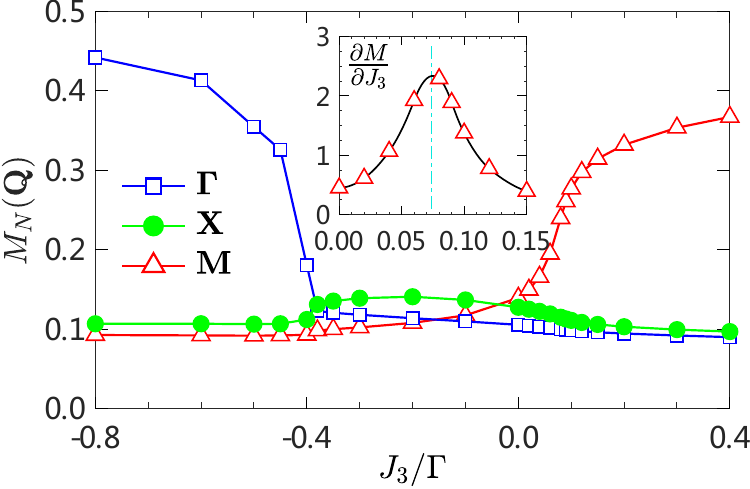}\\
\renewcommand{\figurename}{\textbf{Supplementary Figure}}
\caption{Order parameters $M_N(\textbf{Q})$ for the FM order (blue square), X-correlation (green circle), and zigzag order (red triangle)
with $\textbf{Q}$ = ${\bf{\Gamma}}$, $\textrm{\bf{X}}$, and $\textrm{\bf{M}}$, respectively.
The system size of the XC cylinder is $12\times6$ with a circumference of 6.
Inset: The first derivative of $M_N({\bf{\textrm{\bf{M}}}})$ versus $J_3$.}\label{FIGSM-MSFJ3}
\end{figure}

Fig.~\ref{FIGSM-MSFJ3} shows the evolution of the magnetic orders versus $J_3$ in a rather wide region.
It could be found that the ferromagnetic (FM) and AFM $J_3$ model tend to select the FM phase and zigzag phase, respectively,
as their ground states on the perturbation of AFM $\Gamma$ interaction.
Between the two, the maximum of $M_N(\bf{Q})$ appears at $\textrm{\bf X}$ point
of the Brillouin zone~(see Fig.~\textcolor{red}{3}a of the main text).
At $J_3=0$, $M_N(\bf{Q})$ at $\textrm{\bf M}$ becomes comparable to that at $\textrm{\bf X}$.
However, as can be seen from the inset which shows the first derivative of $M_N({\bf{\textrm{\bf{M}}}})$ versus $J_3$,
the peak locates at a tiny but nonzero $J_{3,t} \approx 0.075$.
This provides further evidence that the ground state of the $\Gamma$ model, in which $J_3$ is zero, is not the zigzag ordering.

\end{document}